\documentclass{style/nseJournal}

\usepackage{isotope}
\usepackage{tikz}
\usepackage{booktabs}
\usepackage{hyperref}

\newbox{\bigpicturebox}

\tikzstyle{nuclide} = [rectangle, rounded corners, minimum width=1cm, minimum height=1cm,text centered, text centered, text width=1.5cm, draw=black]

\tikzstyle{CASLnuc} = [rectangle, rounded corners, minimum width=1cm, minimum height=1cm,text centered, text centered, text width=1.5cm, draw=black, fill=yellow!90]

\tikzstyle{delaynuc} = [rectangle, rounded corners, minimum width=1cm, minimum height=1cm,text centered, text centered, text width=1.5cm, draw=black, fill=gray!40]

\usetikzlibrary{arrows, matrix, positioning, shapes, arrows, decorations.pathreplacing, calc, shapes.geometric, intersections}
\tikzstyle{arrow} = [thick,->,>=stealth]

\begin{document}






\title{Modeling Decay Heat with a \\ Simplified Depletion Chain in OpenMC}

\author{
Tanmay Gupta$^{1,2,}$\thanks{Corresponding author: tg2891@columbia.edu}
\and Benoit Forget$^{2}$
}

\date{
$^1$\,\textit{Department of Physics, Columbia University, New York, NY 10027}\\
$^2$\,\textit{Department of Nuclear Science and Engineering, MIT, Cambridge, MA 02139}
}

\maketitle

\begin{abstract}
OpenMC can be used to computationally model depletion and produce estimates of decay heat. As an input to depletion simulations, OpenMC requires a depletion chain that details nuclide transmutation pathways. The simplified CASL depletion chain was designed to track relatively few nuclides while still accurately modeling the effective neutron multiplication factor and nuclide number densities. However, the CASL chain dramatically underestimates decay heat due to the many nuclides it does not contain. In this work, we modify the CASL depletion chain to improve its accuracy while maintaining its computational efficiency. We demonstrate the effectiveness of adding pseudo-nuclides to the CASL chain, with each pseudo-nuclide capturing the behavior of a large group of nuclides. We further introduce ``delay nuclides,'' which dramatically improve the accuracy of decay heat estimates.
\end{abstract}

\par\vspace{1em}\noindent\textbf{Keywords} --- Decay heat, depletion chain, pseudo-nuclide

\clearpage

\section{Introduction}
\label{sec:introduction}

Nuclear fission produces fission products that are often unstable and undergo radioactive decay. Various transmutation reactions, such as radiative capture, can also create nuclides that will decay. The rate of recoverable energy released by the radioactive decay of such unstable nuclides is called \textit{decay heat}.

In thermal reactors, decay heat accounts for approximately 6 to 7\% of total reactor power~\cite{todreas_kazimi}. For a 3000 MWth reactor, this would account for 180 to 210 MWth, which is quite a significant amount of power. The majority of the energy released from fission is prompt and in the form of kinetic energy of fission products. After shutdown, the fission chain reaction terminates, but the unstable nuclides that have accumulated continue to undergo radioactive decay, making decay heat the dominant source of power post-shutdown.

Thus, heat removal systems must continue removing heat from the core even after shutdown. When such residual heat removal systems fail, the continued production of
decay heat can have severe consequences, as was historically seen in the case of Fukushima. Therefore, accurate modeling of decay heat is essential for design considerations and safety analyses.

Modeling decay heat is a natural part of depletion calculations, but certain simplified decay chains that were developed for core analysis during operation, such as the CASL depletion chain~\cite{Kim_CASL_Specification}, do not properly capture decay heat. In this work, we propose modifications to the simplified CASL depletion chain to more accurately capture decay heat, and we verify the modified chain's performance using the OpenMC radiation transport code.

\section{Background}
\label{sec:background}
OpenMC is an open-source Monte Carlo transport code that, in addition to modeling neutron and photon transport, is also capable of modeling the process of depletion, or the change in nuclide number densities over time as nuclides undergo transmutation reactions and radioactive decays~\cite{romano_openmc_2015}. 
In order to solve for the nuclide number densities $N_i(t),$ OpenMC uses various numerical methods to solve
\begin{equation} \label{eq:NNDs}
  \begin{cases} \dfrac{d\boldsymbol{n}}{dt} = \boldsymbol{A}(\boldsymbol{n}, t) \, \boldsymbol{n} \\ 
  \boldsymbol{n}(0) = \boldsymbol{n_0} \end{cases},
\end{equation}
where $\boldsymbol{n}(t)$ is the vector containing all nuclide number densities $N_i(t)$, $\boldsymbol{A}$ is the burnup matrix with transmutation reaction rates and decay coefficients, and $\boldsymbol{n_0}$ is the vector with initial nuclide number densities, as specified by the user.

From the calculated nuclide number densities $N_i(t)$, total decay heat can be obtained as a function of time via 
\begin{equation} \label{eq:DecayHeat}
  P_{decay} (t) = \sum_{i} \lambda_i Q_i N_i(t),
\end{equation}
where $\lambda_i$ is the decay constant, and $Q_i$ is the average recoverable energy released per decay. Note that throughout this work, consistent with the ENDF format~\cite{ENDF_6_Format_brown_endf-6_2023}, $Q_i$ excludes the energy from neutrinos released during decays.

In order to construct the burnup matrix $\boldsymbol{A}$ used to calculate nuclide number densities, OpenMC requires information about possible transmutation reactions and decay pathways (e.g., branching ratios, fission product yields), which is all contained within a \textit{depletion chain} file. OpenMC is capable of running depletion calculations with different depletion chains. One such chain, called the \textit{ENDF depletion chain} throughout this work, is based on data from the ENDF/B-VII.1 library~\cite{ENDF}. The ENDF depletion chain contains data for the reactions and decay pathways of over 3,800 nuclides, serving as the benchmark for our decay heat calculations. 

Nevertheless, running depletion analyses with such a large depletion chain is computationally expensive. In the process of numerically integrating Eq.~\eqref{eq:NNDs}, OpenMC evaluates matrix exponentials based on the Chebyshev Rational Approximation Method (CRAM). Previous work has shown that CRAM runtime scales approximately linearly with the number of nuclides tracked~\cite{Calvin_PNs}. Other depletion solvers are also reported to show runtime increasing with the number of reactions and decays in the matrix~\cite{Wieselquist_SCALE_Depletion}.

Furthermore, the memory requirements of depletion scale linearly with the number of nuclides. In each distinct depletable region, OpenMC must store the density of each nuclide in the depletion chain. Models of commercial light water reactors can have hundreds of millions of depletable regions~\cite{Griesheimer_Large_Scale_MC}, each with its own reaction rates and nuclide number densities. Simply storing the densities and reaction rates will require substantial memory usage, requiring hundreds or thousands of gigabytes of memory~\cite{Forget_Yu_Depletion}. 

Additionally, in transport-coupled depletion simulations---used in cases when the neutron flux is dependent on the nuclide densities---OpenMC has to find a solution to the transport equation to determine reaction rates for each timestep. For a Monte Carlo code, such as OpenMC, the transport solver time is much higher than the depletion solver time~\cite{romano_depletion_2021}. Using the ENDF depletion chain requires tracking a large number of reaction rate tallies, which drastically increases transport runtime. Therefore, if reducing runtime for OpenMC was the main goal, a depletion chain with a full list of nuclides but a simplified set of reactions would suffice. However, reducing the number of nuclides tracked would still reduce the enormous memory requirements, as well as further decrease depletion solver times, which can be more significant for deterministic codes.

In order to reduce the aforementioned computational costs of using a detailed depletion chain, researchers typically develop simplified chains, as was done at Oak Ridge National Laboratory~\cite{Wieselquist_SCALE_Depletion}. OpenMC offers a depletion chain containing only 228 nuclides based on specifications of this simplified depletion chain~\cite{Kim_CASL_Specification}. As the prior work was associated with the Consortium for Advanced Simulation of Light Water Reactors (CASL), we henceforth refer to this chain as the \textit{CASL depletion chain}. The CASL chain was designed to preserve the effective multiplication factor and number densities of tracked nuclides that impact reactivity during operation~\cite{romano_depletion_2021}. Furthermore, due to the computational benefits of the CASL chain, it is recommended for ``general use'' by OpenMC users~\cite{romano_depletion_2021}.

However, the CASL chain drastically underestimates decay heat. This work aims to increase the accuracy of the CASL chain's decay heat estimates while maintaining its computational efficiency.

\section{Methods}
\label{sec:methods}

\subsection{Depletion Models}
\label{subsec:depletion model}

\subsubsection{PWR Depletion Model}
\label{subsubsec:pwr depletion model}
Our primary depletion model was a pressurized water reactor (PWR) fuel pin cell with reflective boundaries, equivalent to an infinite lattice of PWR pin cells. The fuel used was 4.25\% enriched uranium dioxide. The geometry and all other material compositions at the beginning of cycle were based on the fuel pin in the BEAVRS benchmark~\cite{BEAVRS}. A heating rate of 174 W/cm was specified, in accordance with previous models~\cite{romano_depletion_2021}.

\subsubsection{SFR Depletion Model}
\label{subsubsec:sfr depletion model}
To further test our modified depletion chains, we ran depletion simulations for a sodium-cooled fast reactor (SFR) problem as well. Our model is based on the SFR assembly model used by Romano et al.~\cite{romano_depletion_2021}, but we utilize a single hexagonal fuel pin cell with periodic boundary conditions, again equivalent to an infinite lattice of pin cells. The geometry and material compositions were based on specifications for an outer core fuel pin of a 1000 MWth medium-size metallic core (MET-1000), provided in a benchmark established by the OECD's Nuclear Energy Agency~\cite{OECD_NEA}. The heating rate was set to 233 W/cm.

\subsubsection{Timesteps}
\label{subsubsec:timesteps}
In both the PWR and SFR models, the pin cells were depleted for approximately 4.5 years at full power. Timesteps were initially uneven, with smaller timesteps at the beginning in order to accurately build up equilibrium nuclides. The pin cells were then further depleted with a source rate of zero, resulting in decay-only timesteps, representing depletion post-shutdown. For depletion, we used the second-order CE/CM predictor-corrector algorithm available in OpenMC. 

\subsubsection{Execution Settings}
\label{subsubsec:execution settings}
For all simulations, 20 inactive batches, 80 active batches, and 500,000 particles per batch were used. This is consistent with the execution parameters used in the work of Yu and Forget~\cite{Forget_Yu_Depletion}. Simulations were run on a single 192-core node (dual socket AMD EPYC 9654) on the MIT Office of Research Computing and Data's Engaging Cluster.

\subsection{Underestimation of Decay Heat}
\label{subsec:underestimation of decay heat}

\begin{figure}[h!]
\centering
\begin{subfigure}[h]{0.475\textwidth}
   \includegraphics[width=1\linewidth]{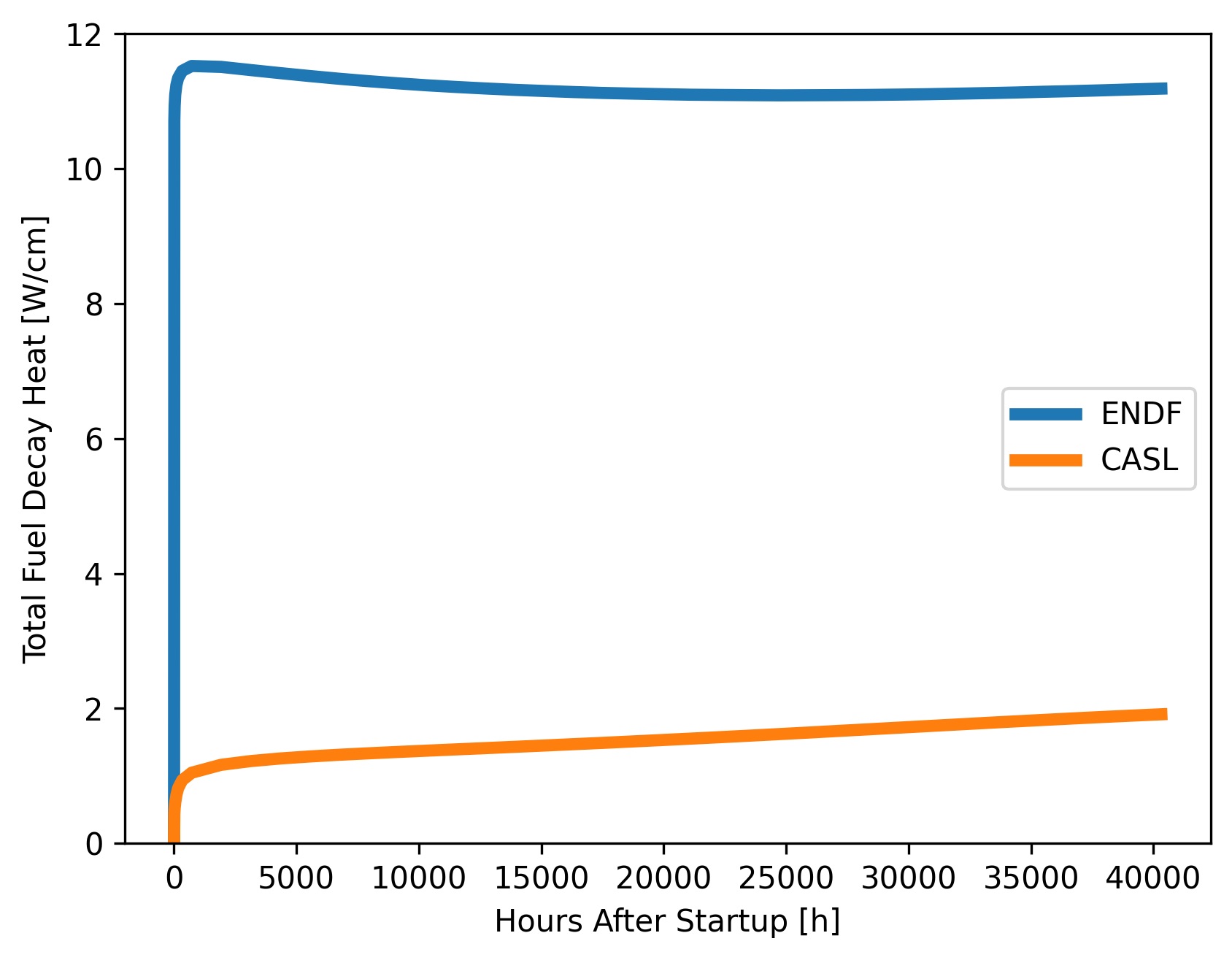}
   \caption{Operation}
   \label{subfig:The Problem - Operation} 
\end{subfigure}
\begin{subfigure}[h]{0.475\textwidth}
   \includegraphics[width=1\linewidth]{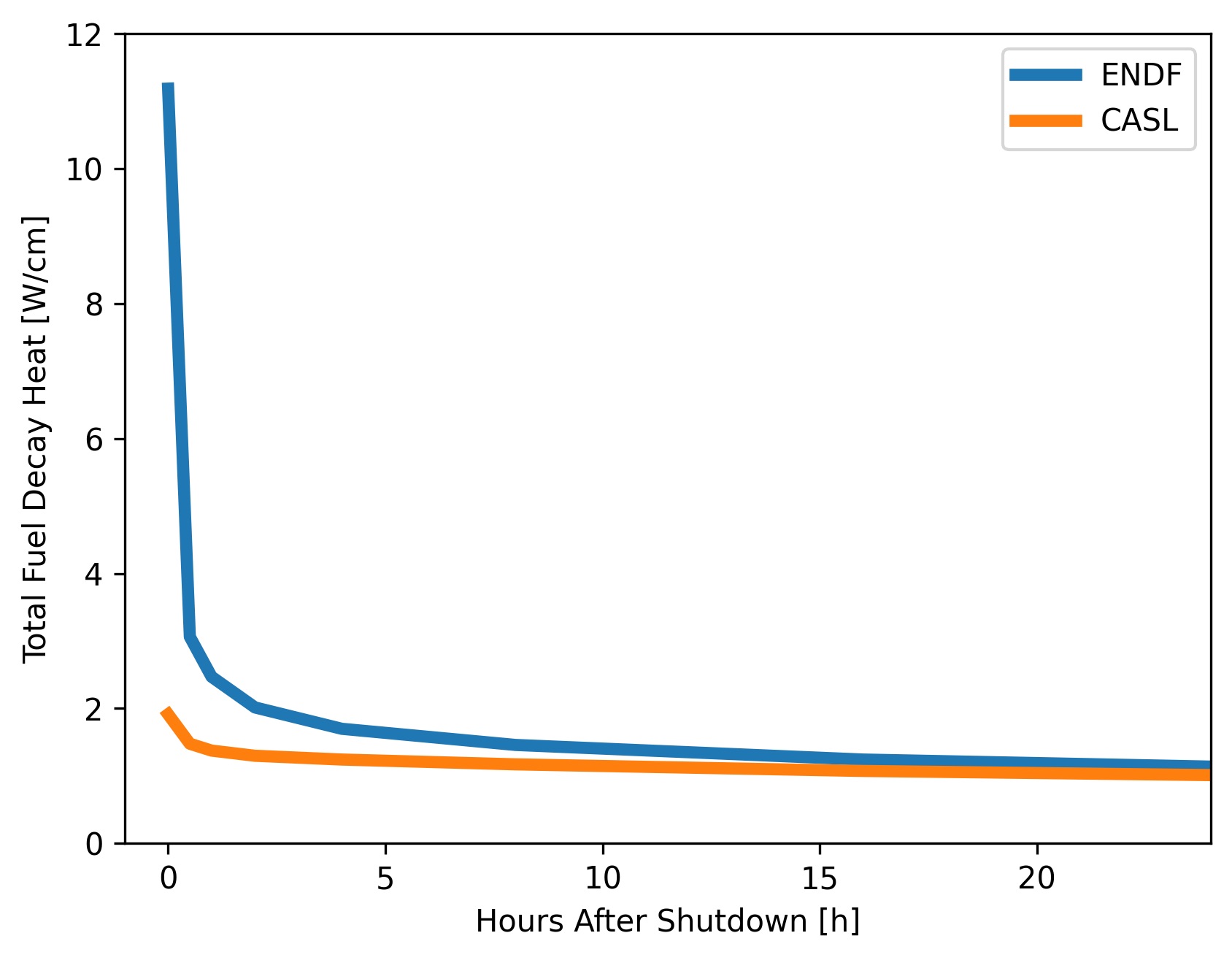}
   \caption{Post shutdown}
   \label{subfig:The Problem - Post-shutdown}
\end{subfigure}
\caption{\protect\subref{subfig:The Problem - Operation} CASL chain underestimates decay heat during operation. \protect\subref{subfig:The Problem - Post-shutdown} After shutdown, underestimation persists.}
\end{figure}

Initially using the PWR pin cell model, we ran depletion simulations using both the ENDF chain and the CASL chain. Using Eq.~\eqref{eq:DecayHeat}, we were able to obtain decay heat from the nuclide densities that OpenMC calculated for each time step. As shown in Figure~\ref{subfig:The Problem - Operation}, the ENDF chain's decay heat estimates during operation fall within 6\% to 7\% of the specified heating rate of 174~W/cm as expected. Compared to the ENDF benchmark, however, the CASL chain drastically underestimates decay heat throughout the entire operation time interval. Figure~\ref{subfig:The Problem - Post-shutdown} demonstrates that the CASL chain continues to underestimate decay heat post-shutdown.

\subsection{Characterizing the Problem}
\label{subsec:characterizing the problem}

We first aimed to understand why the CASL chain underestimates decay heat. As the CASL chain's nuclides are a subset of the full ENDF chain's nuclides, we can split the ENDF nuclides into \textit{ENDF-exclusive nuclides}, and the nuclides that it shares with the CASL chain, called \textit{shared nuclides}.

\begin{figure}[ht]
  \centering
  \includegraphics[width=70mm]{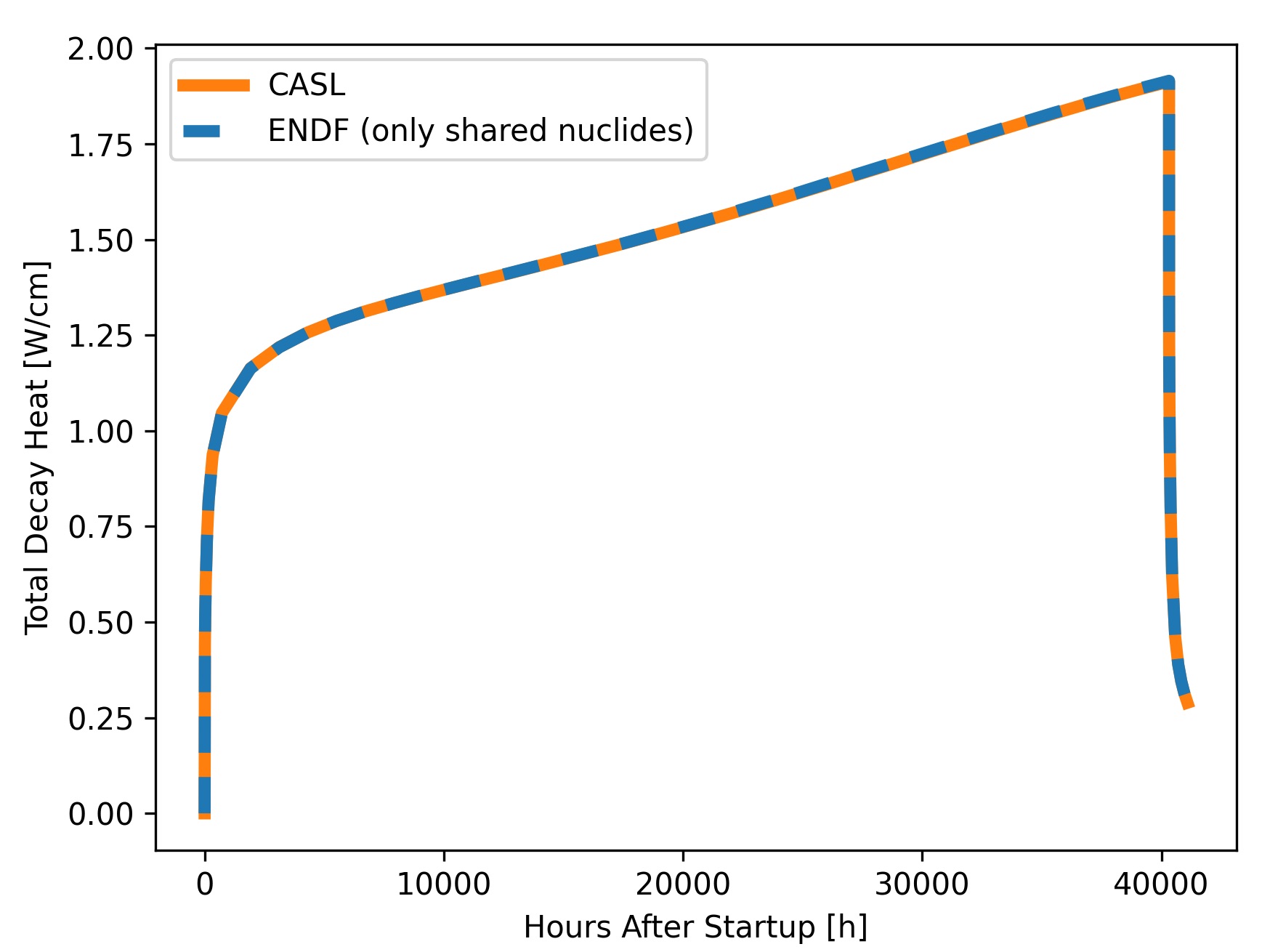}
  \caption{Total decay heat contribution of nuclides shared between ENDF and CASL in depletion simulations for each of the two chains.}
  \label{fig:SharedNuclidesMatch}
\end{figure}

Figure \ref{fig:SharedNuclidesMatch} displays the decay heat calculated by the CASL chain, as well as the sum of the contributions of the shared nuclides in the results from using the ENDF chain. These two quantities match throughout operation as well as post-shutdown. This is reasonable as the CASL chain accurately models the densities of the nuclides it tracks, which are the shared nuclides. Thus, the simplified chain's underestimation is due to the decay heat contributions of the ENDF-exclusive nuclides, which are not tracked in the simplified depletion chain.

Next, we aimed to determine if only a few ENDF-exclusive nuclides accounted for this difference. This would allow us to add those nuclides to the CASL chain, improving accuracy without tracking a large number of nuclides. However, after conducting an analysis of the ENDF-exclusive nuclides, we observed that hundreds of nuclides would have to be added to the CASL chain to even approach the decay heat of the ENDF chain.

\subsection{Pseudo-Nuclides}
\label{subsec:pseudo-nuclides}
To capture the decay heat of hundreds of nuclides without actually tracking each individual nuclide, we used \textit{pseudo-nuclides}. A pseudo-nuclide (PN) is a fictitious nuclide that captures a desired property, in this case decay heat, for a group of many different nuclides. By adding PNs to the CASL depletion chain, they accumulate and contribute to total decay heat. The goal is to have the decay heat of the PN approximate the sum of decay heat contributions of all the nuclides it represents. The idea of using fictitious nuclides to approximate the behavior of many nuclides was used early on for reactivity analysis~\cite{aldama2003wims}, and is similar in concept to commonly used delayed neutron groups. Similar approaches have been used for decay heat, with some authors using ``decay heat precursors'' (DHPs)~\cite{Huang_DHPs} or ``pseudo-decay nuclides''~\cite{ZHANG2026104066} whose decay constants and fission yields are determined through an optimization process. On the other hand, the prior work of the authors, which is extended here, is based on grouping nuclides of similar decay constants~\cite{Gupta_Forget_ANS_Conference}. This more physics-based approach follows the work of Calvin~\cite{Calvin_PNs} in using PNs that can be produced as fission, reaction, or decay products, whereas DHPs are only produced as fission products. 

\begin{figure}[ht]
  \centering
  \includegraphics[width=70mm]{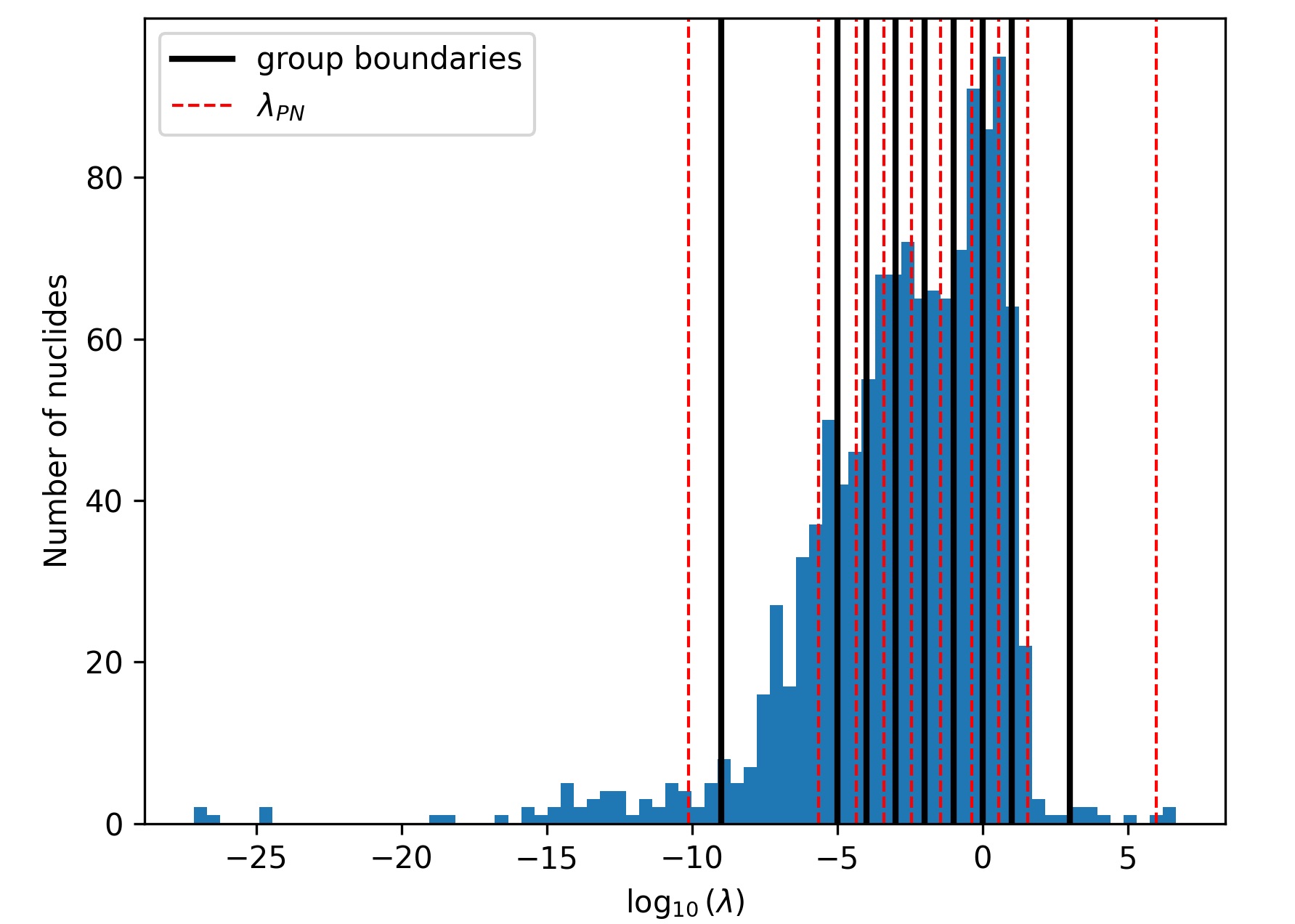}
  \caption{Nuclides are grouped into ten groups based on decay constant. Each group is represented by a single PN with decay constant $\lambda_{PN}$.}
  \label{fig:10_PNs}
\end{figure}

Figure~\ref{fig:10_PNs} shows a histogram of the ENDF-exclusive nuclides whose decay heat we want to capture (nuclides that are not possible to reach from the initial composition were excluded). The solid black lines delineate the boundaries of the groups, each containing nuclides with similar decay constants. Note that nuclides were apportioned into groups that were similarly sized on the scale of $\log_{10}(\lambda_i)$ as this results in a more even distribution of decay constants into groups. If the groups were similarly sized on a linear scale, then all of the nuclides with small decay constants would end up in a single group.

Furthermore, a representative decay constant $\lambda_{PN}$, shown by a red dotted line, was chosen by taking the average of the decay constants of nuclides in the group. We chose to use 10 PNs as it results in relatively few additional nuclides to be tracked, while creating a marked improvement in decay heat accuracy.

\subsubsection{Theory of PNs}
\label{subsubsec:theory of PNs}
The exact decay heat for each group is given by Eq.~\eqref{eq:DecayHeat}. However, within a group, all decay constants are approximately equal to the representative decay constant $\lambda_{PN}.$
Thus, within each group, we have 
\begin{equation} \label{eq: lambda_PN approximation}
    \sum_i \lambda_i Q_i N_i \approx \sum_i \lambda_{PN} Q_i N_i = \lambda_{PN} \sum_i Q_i N_i.
\end{equation}
Hence, if we can get OpenMC to track the quantity $\sum_i Q_i N_i$ for each group, we solely need to multiply by $\lambda_{PN}$ at the end to obtain decay heat.

For each nuclide, the differential equation governing its density is given by
\begin{equation}\label{eq: dn/dt prod-decay}
    \frac{dN_i}{dt} = P_i(t) - \lambda_i N_i,
\end{equation}
where $P_i(t)$ is the production rate of the nuclide. Note that we assume that any reaction for these ENDF-exclusive nuclides is of negligible importance; thus removal consists only of a decay term.

Within a group, we make the approximation $\lambda_i \approx \lambda_{PN}.$ We substitute into Eq.~\eqref{eq: dn/dt prod-decay} and then multiply through by $Q_i$, giving
\begin{equation}\label{eq: d(qn)/dt prod-decay}
    \frac{d(Q_iN_i)}{dt} = Q_iP_i(t) - \lambda_{PN} Q_iN_i.
\end{equation}
Summing over all nuclides in the group gives
\begin{equation}\label{eq: sum Q_iN_i}
    \dfrac{d ( \sum_i Q_i N_i  ) }{dt} = \left( \sum_i Q_i P_i(t) \right) - \lambda_{PN} \left( \sum_i  Q_i N_i \right).
\end{equation}
From this equation, we can deduce that if a pseudo-nuclide with decay constant $\lambda_{PN}$ is produced at a rate of $\sum_i Q_i P_i(t),$ the ``number density'' of this PN will be $\sum_i Q_i N_i.$ This is precisely the quantity we needed to track to obtain the group's total decay heat, as shown in Eq.~\eqref{eq: lambda_PN approximation}.

This method of multiplying production rates by $Q_i$ has been used in prior work~\cite{Calvin_PNs}. One could multiply production rates by $Q_i \lambda_i$ as well, and then the number density of the PN would itself be the total decay heat of the group. However, we note that multiplying only by $Q_i$ has the benefit that the decay heat is extremely accurate in equilibrium situations. From Eq.~\eqref{eq: dn/dt prod-decay}, at equilibrium, $\lambda_iN_i = P_i$, so the decay heat $\lambda_iN_iQ_i$ equals $P_iQ_i$. Looking at Eq.~\eqref{eq: sum Q_iN_i}, if the left-hand side is small, then $\lambda_{PN} \left( \sum_i  Q_i N_i \right)$, which is the decay heat of the PN, equals $\left( \sum_i Q_i P_i(t) \right).$ In this case, there is much less error from the approximation of $\lambda_i \approx \lambda_{PN}.$ Also, note that in the initial fuel material composition, we assumed no ENDF-exclusive nuclides that would contribute to decay heat. If there are unstable nuclides present at the beginning, then their initial number densities would need to be scaled by $Q_i$ and added to the appropriate PN.

\subsubsection{Implementation of PNs}
\label{subsubsec:implementation of PNs}

Setting the decay constant of the PN to the appropriate $\lambda_{PN}$ is relatively simple to do when creating the nuclide in OpenMC. However, producing the PN at a rate of $\sum_i Q_i P_i(t)$ requires more careful consideration, as described below.

The number densities of the CASL nuclides are accurately captured by the CASL chain, and the ENDF-exclusive nuclides are produced from the CASL nuclides. Thus, the fundamental idea in our implementation is the following: iterate through each CASL nuclide, and for each pathway to produce an ENDF-exclusive nuclide, add a production pathway for the appropriate PN to the CASL nuclide within the CASL chain. We can ascertain the appropriate yields and branching ratios from the ENDF chain and also scale these by $Q_i$ to obtain the desired production rate.

For nuclides that are one ``degree'' away from a CASL nuclide (fission, reaction, or decay product), we can do this process exactly. However, within the ENDF chain, a CASL nuclide can also produce ENDF-exclusive nuclides indirectly through decay pathways.

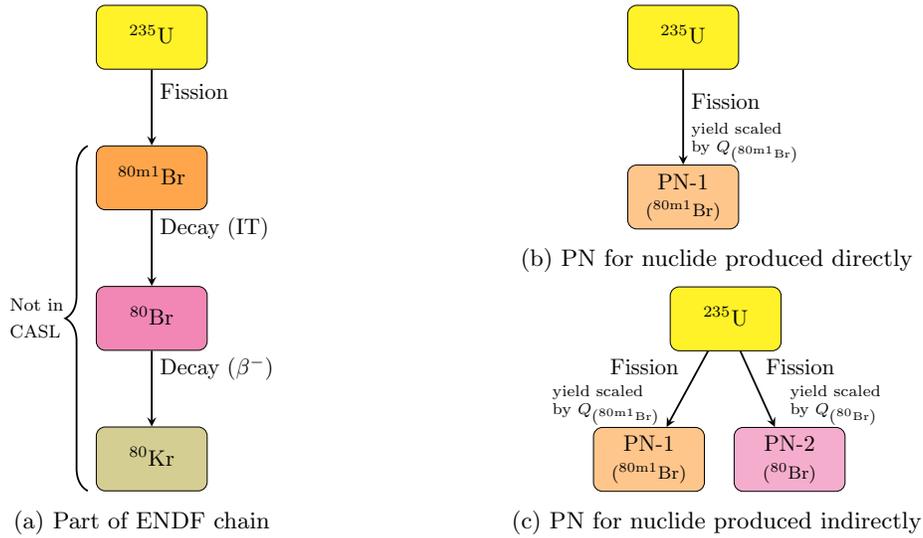
\begin{figure}[h]
\centering
\sbox{\bigpicturebox}{%
  \begin{subfigure}[b]{.4\columnwidth}
    \centering
    \scalebox{0.85} {
    \begin{tikzpicture}[scale=1][node distance=1.0cm]
      \node (U235) [CASLnuc, xshift=1cm, yshift=4cm] {\isotope[235]{U}};
      \node (Br80_m1) [nuclide, fill=orange!70, below of=U235, xshift=0cm, yshift=-1.2cm] {\isotope[80\text{m1}]{Br}};

      \node (Br80) [nuclide, fill=magenta!60, below of=Br80_m1, xshift=0cm, yshift=-1.2cm] {\isotope[80]{Br}};

      \node (Kr80) [nuclide, fill=olive!40, below of=Br80, xshift=0cm, yshift=-1.2cm] {\isotope[80]{Kr}};

      \draw [arrow] (U235) -- (Br80_m1) node[midway,above right] {Fission};
      \draw [arrow] (Br80_m1) -- (Br80) node[midway,above right] {Decay (IT)};
      \draw [arrow] (Br80) -- (Kr80) node[midway,above right] {Decay ($\beta^-$)};

      \draw [decorate,thick,decoration={brace, amplitude=10pt, mirror}] (0, 2.3) -- (0, -3.1) node [black,midway,xshift=-0.8cm,align=left] {\footnotesize Not in\\\footnotesize CASL};
    \end{tikzpicture}
    }
  \caption{Part of ENDF chain}
  \label{subfig: part of ENDF chain}
  \end{subfigure}
}
\usebox{\bigpicturebox}\hfill
\begin{minipage}[b][\ht\bigpicturebox][s]{.58\columnwidth}
\begin{subfigure}{\textwidth}
  \centering
  \scalebox{0.85} {
  \begin{tikzpicture}[scale=1][node distance=1.2cm]
  \node (U235) [CASLnuc, xshift=0cm, yshift=4cm] {\isotope[235]{U}};
  \node (Br80_m1) [nuclide, fill=orange!45, below of=U235, xshift=0cm, yshift=-1.5cm] {PN-1 \footnotesize(\isotope[80\text{m1}]{Br})};

  \draw [arrow] (U235) -- (Br80_m1) node[name=Fission1,midway,above right] {Fission};
  \node [below of=Fission1, xshift=0.3cm, yshift=0.35cm, align=left]{\scriptsize yield scaled\\[-0.4em]\scriptsize by $Q_{\left(\isotope[80\text{m1}]{Br}\right)}$};
  \end{tikzpicture}
  }
  \caption{PN for nuclide produced directly}
  \label{subfig: PN for nuclide produced directly}
\end{subfigure}
\vfill
\begin{subfigure}[b]{\textwidth}
  \centering
  \scalebox{0.85}{
  \begin{tikzpicture}[scale=1][node distance=1.2cm]
    \node (U235) [CASLnuc, xshift=0cm, yshift=4cm] {\isotope[235]{U}};
    \node (Br80_m1) [nuclide, fill=orange!45, below of=U235, xshift=-1.2cm, yshift=-1.2cm] {PN-1 \footnotesize(\isotope[80\text{m1}]{Br})};

    \node (Br80) [nuclide, fill=magenta!40, right of=Br80_m1, xshift=1.2cm, yshift=0cm] {PN-2 \footnotesize(\isotope[80]{Br})};

    \draw [arrow] (U235) -- (Br80_m1) node[name=Fission1,midway,above left, yshift=0.1cm] {Fission};
    \draw [arrow] (U235) -- (Br80) node[name=Fission2,midway,above right, yshift=0.1cm] {Fission};

    \node [below of=Fission1, xshift=-0.6cm, yshift=0.4cm,align=left] {\scriptsize yield scaled\\[-0.4em]\scriptsize by $Q_{\left(\isotope[80\text{m1}]{Br}\right)}$};
    \node [below of=Fission2, xshift=0.55cm, yshift=0.4cm, align=left] {\scriptsize yield scaled\\ [-0.4em]\scriptsize by $Q_{\left(\isotope[80]{Br}\right)}$};
  \end{tikzpicture}
  }
\caption{PN for nuclide produced indirectly}
\label{subfig: PN for nuclide produced indirectly}
\end{subfigure}
\end{minipage}
\caption{Process of Adding PNs to CASL Chain}
\label{fig: Adding PNs Example}
\end{figure}

Consider the example shown in Figure~\ref{fig: Adding PNs Example}. In the small portion of the ENDF depletion chain shown in Figure~\ref{subfig: part of ENDF chain}, \isotope[235]{U} produces \isotope[80\text{m1}]{Br} via fission, which subsequently undergoes an isomeric transition to \isotope[80]{Br}. As \isotope[80]{Br} is also unstable, it can undergo a $\beta^-$ decay into \isotope[80]{Kr}.

While \isotope[235]{U} is in the CASL chain, the remaining nuclides are not. Thus, the decay heat from the IT and $\beta^-$ decays are not captured by the CASL chain.

In order to capture the decay heat from the decay of \isotope[80\text{m1}]{Br}, in the CASL chain, we can add the appropriate PN as a fission product (say \isotope[80\text{m1}]{Br} belongs to the group represented by PN-1). The fission yield for PN-1 is taken as the fission yield for \isotope[80\text{m1}]{Br} from the ENDF chain multiplied by its average decay energy. Figure~\ref{subfig: PN for nuclide produced directly} demonstrates this initial step. The decay heat contribution from this production of PN-1  accounts for the IT decay heat.

However, it remains to capture the decay heat of the $\beta^-$ decay. Suppose PN-2 is the PN representing \isotope[80]{Br}. We do not have PN-1 produce PN-2 because PN-1 represents more nuclides than just \isotope[80\text{m1}]{Br}, and not all of those nuclides decay to a nuclide that is represented by PN-2. Furthermore, as the material gets depleted, the ratios of number densities of each of the nuclides represented by a single PN continue to change. Thus, having a PN decay to other PNs with fixed branching ratios would not be accurate either. 

Thus, we initially adopt the approach based on the prior work of Calvin~\cite{Calvin_PNs}. We approximate the decay to occur instantaneously, having the CASL nuclide produce PNs to capture decay heat from decay products. Returning to the example, as shown in Figure~\ref{subfig: PN for nuclide produced indirectly}, we produce PN-2 from \isotope[235]{U}. This process is a good approximation only if the parent nuclide is relatively short-lived. If our requirement for short-lived is too strict, we risk underestimating decay heat. On the other hand, if we treat long-lived nuclides as if they decay instantaneously, the production of decay products would not be physical. After experimentation with different cutoffs, we have chosen to apply this instantaneous decay only if the half-life of the parent nuclide is less than a week.

In the example in Figure~\ref{fig: Adding PNs Example}, we considered a simple portion of the ENDF depletion chain that did not have any branches. In the full algorithm, branching ratios are tracked as the ENDF chain is traversed. The production of PNs is appropriately scaled by a factor capturing the product of all the appropriate branching ratios. For instance, in our example, the production of PN-1 would be scaled by the branching ratio of the IT. Also, the production of PN-2 would be scaled by the branching ratio of the IT and the branching ratio of the $\beta^-$ decay. 

\section{Results for 10 Pseudo-Nuclides}
\label{sec:results for 10 pseudo-nuclides}

\begin{figure}[h!]
\centering
\begin{subfigure}[h]{0.475\textwidth}
   \includegraphics[width=1\linewidth]{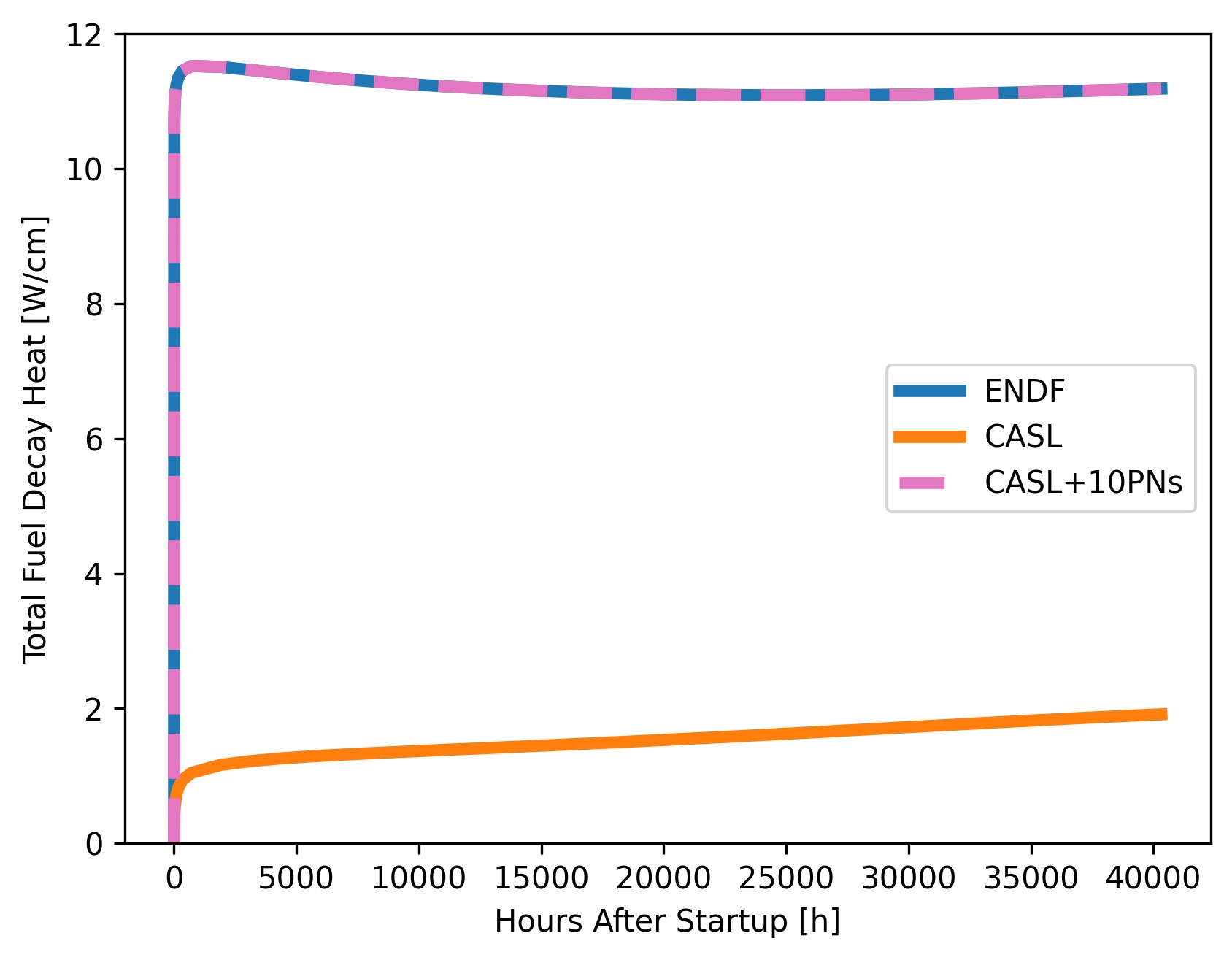}
   \caption{Operation}
   \label{subfig:10PNs_Operation} 
\end{subfigure}
\begin{subfigure}[h]{0.475\textwidth}
   \includegraphics[width=1\linewidth]{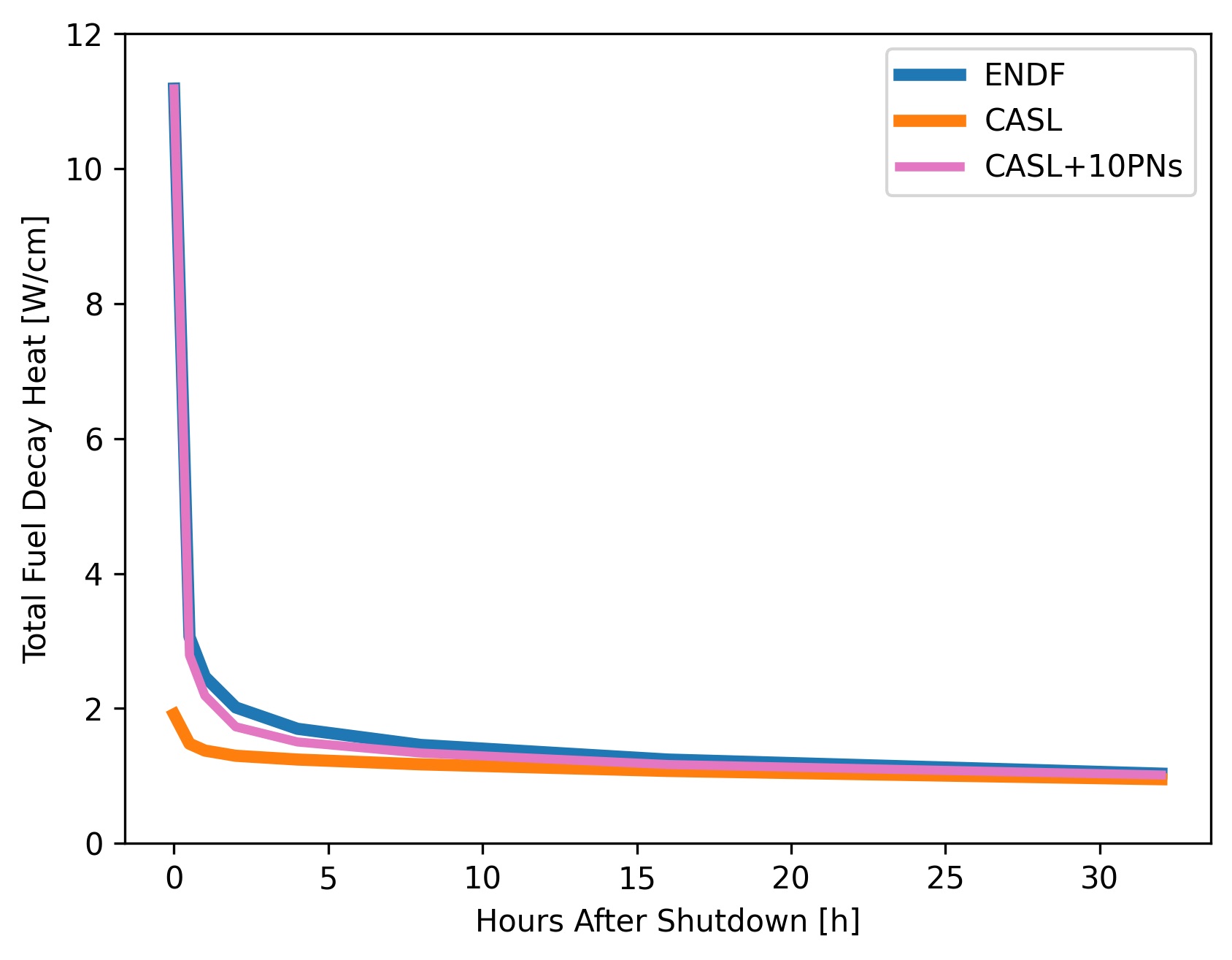}
   \caption{Post shutdown}
   \label{subfig:10PNs_Post_Shutdown}
\end{subfigure}
\caption{Adding 10 PNs to the CASL chain improves decay heat estimates both \protect\subref{subfig:10PNs_Operation} during operation and \protect\subref{subfig:10PNs_Post_Shutdown} post shutdown.}
\end{figure}

Figure~\ref{subfig:10PNs_Operation} demonstrates the decay heat calculated by the CASL chain with 10 PNs added (CASL+10PNs) during operation. This depletion chain matches the ENDF benchmark decay heat calculations much more closely than the CASL chain. Figure~\ref{subfig:10PNs_Post_Shutdown} demonstrates the improvement in decay heat calculations post-shutdown by adding the PNs.

Furthermore, we observed that adding 10 PNs to the CASL chain maintained the runtime benefits of this simplified chain. When including the transport solve, both the CASL and CASL+10PNs chains took approximately 50\% less total runtime than the ENDF chain. This was likely due to the fewer reactions in these chains, as tallying reactions during transport tends to dominate runtime for MC codes. When considering the average depletion solver time (not including transport time) during operation, the CASL chain ran 71\% faster than the ENDF chain. The CASL+10PNs chain maintained this computational advantage, running 70\% faster than the ENDF chain. After shutdown, during the decay-only steps, both the CASL and the CASL+10PNs chains ran approximately 13\% faster than the ENDF chain.

\begin{figure}[ht]
  \centering
  \includegraphics[width=70mm]{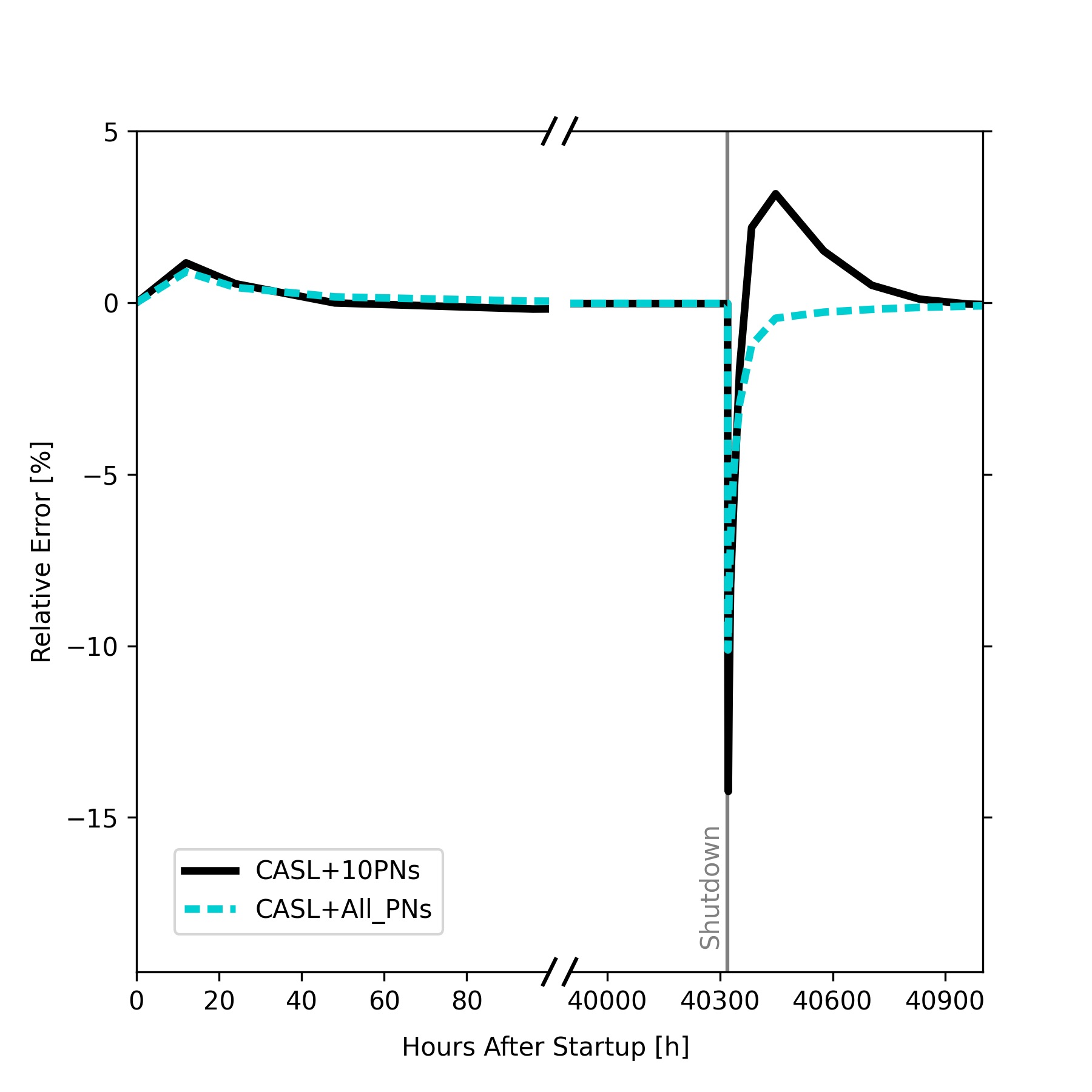}
  \caption{CASL+10PNs and CASL+All\_PNs have spikes in relative error at startup and shutdown.}
  \label{fig:Relative_Error}
\end{figure}

Starting with a simplified depletion chain not designed to preserve decay heat, we were able to mostly recover the ENDF benchmark decay heat estimates with the addition of 10 pseudo-nuclides. These results show that 10 PNs were able to capture the decay heat behavior of hundreds of nuclides. This was accomplished while maintaining the relatively low number of nuclides tracked by the CASL chain, preserving the associated computational benefits in terms of both runtime and memory usage.

\section{Delay Nuclides}
\label{sec:delay nuclides}

Figure~\ref{fig:Relative_Error} shows the relative error in the CASL+10PNs decay heat when compared to the ENDF benchmark decay heat. An examination reveals that the largest errors occur when the system is not at equilibrium. The CASL+10PNs chain's calculated decay heat has appreciable deviation from the ENDF chain in the early hours following startup and shutdown.

To eliminate any error from how nuclides were grouped, we conducted an analysis where instead of using 10 PNs, all ENDF-exclusive nuclides got their own PN. With this \textit{CASL+All\_PNs} chain, each nuclide is represented by a PN with its exact decay constant, with no approximations made. In other words, there are as many PN groups as ENDF-exclusive nuclides. Yet, the startup and shutdown errors remain, as shown in Figure~\ref{fig:Relative_Error}.

This indicates that the error comes from the production rates of the PNs. In fact, the spikes at startup and shutdown are an artifact of the loss of information about the structure of decay pathways when using PNs as in the implementation described previously.

\subsection{Instantaneous Decay Causes Startup and Shutdown Error}
\label{subsec:instantaneous decay causes startup and shutdown error}

The error in production rates can be traced back to the instantaneous decay in the implementation of the PNs. Consider once again the example in Figure~\ref{fig: Adding PNs Example}. In the modified CASL chain, after shutdown, no fission will occur. Thus, production of PN-2 will stop. However, PN-2 represents \isotope[80]{Br}, which should continue to be produced from the decay of \isotope[80\text{m1}]{Br}. This leads to the underestimation of decay heat post shutdown. During startup, the production of PN-2 occurs too rapidly since it does not have to go through a decay. Thus, the issue is that the production rates of PNs is not necessarily accurate in time---production is accelerated due to the instantaneous decay.

By comparing the decay heat contributions by nuclide for the CASL+All\_PNs chain and the ENDF chain, we can find the nuclides with the largest error during startup and shutdown. During shutdown, the three nuclides with the largest errors were \isotope[134]{I}, \isotope[138]{Cs}, and \isotope[97\text{m1}]{Nb}. Looking at the production pathways of these nuclides, we can notice a pattern: in the decay chain to produce each of these nuclides, there is a nuclide with relatively long half-life. Thus, the instantaneous decay approximation is not the most accurate. For \isotope[134]{I}, its parent nuclide is \isotope[134]{Te}, which has a half-life around 42 minutes. For \isotope[138]{Cs}, its parent nuclide is \isotope[138]{Xe}, which has a half-life of approximately 14 minutes. For \isotope[97\text{m1}]{Nb} (and \isotope[97]{Nb}), both can be produced through the decay of \isotope[97]{Zr}, which has a half-life of nearly 17 hours.

Note that this effect is also dependent on the size of the timesteps for startup and shutdown. In our model, startup had an initial timestep of 12 hours, whereas post-shutdown timesteps started with 0.5 hours. This explains why \isotope[138]{Cs} and \isotope[134]{I} have large errors after shutdown, but not during startup. The delay in production during startup is small relative to 12 hours, so the instantaneous decay approximation is reasonable, but highly dependent on the time step size. On the other hand, \isotope[97\text{m1}]{Nb} has a delay in production comparable to the scale of 12 hours, so it also has large error from instantaneous decay during startup too.

\subsection{Introducing Delay Nuclides to Improve Production Rates}
\label{subsec:introducing delay nuclides to improve production rates}

In order to improve the accuracy of the production rates, and consequently the accuracy of decay heat estimates, we introduce \textit{delay nuclides}. Their sole purpose is to delay the production of PNs, avoiding the issues that arise from instantaneous decay.

\begin{figure}[h]
  \sbox{\bigpicturebox}{%
  \begin{subfigure}[b]{.5\columnwidth}
  \scalebox{0.85} {
    \begin{tikzpicture}
      \node (CASL) [CASLnuc, xshift=1cm, yshift=4cm] {CASL Nuclide};
      \node (decay_start) [nuclide, fill=cyan!40, below of=CASL, xshift=-1.75cm, yshift=-0.5cm] {};
      \node (dots) [below of=decay_start,yshift=-0.35cm,rotate=90] {$\boldsymbol{\cdots}$};

      \node (decay_end) [nuclide, fill=lime!50, below of=dots, xshift=0cm, yshift=-0.35cm] {};

      \node (A) [nuclide, fill=violet!50, below of=decay_end, xshift=2.0cm, yshift=-0.5cm] {A};

      \draw [arrow] (CASL) -- (decay_start);
      \draw [arrow] (decay_start) -- (dots);
      \draw [arrow] (dots) -- (decay_end);
      \draw [arrow] (decay_end) -- (A);
      \draw [thick,->,>=stealth,dashed] (CASL) -- (A);

      \draw [decorate,thick,decoration={brace, amplitude=10pt, mirror}] (-1.6, 3.1) -- (-1.6, -0.9) node [black,midway,xshift=-1cm,align=left] {\footnotesize Decay \\\footnotesize Pathway};
    \end{tikzpicture}
  }
  \caption{ENDF chain}
  \label{subfig: ENDF delay nucs}
  \end{subfigure}
  }
  \usebox{\bigpicturebox}\hfill
  \begin{subfigure}[b][\ht\bigpicturebox]{.5\columnwidth}
  \centering
  \scalebox{0.85} {
    \begin{tikzpicture}
      \node (CASL) [CASLnuc, xshift=1cm, yshift=4cm] {CASL Nuclide};
      \node (delay) [delaynuc, below of=CASL, xshift=-1.5cm, yshift=-1.85cm] {Delay};

      \node (PN-A) [nuclide, fill=violet!30, below of=delay, xshift=2.5cm, yshift=-1.85cm] {PN-A};

      \draw [arrow] (CASL) -- (delay);
      \draw [arrow] (delay) -- (PN-A);
      \draw [thick,->,>=stealth,dashed] (CASL) -- (PN-A);
    \end{tikzpicture}
  }
  
  \caption{Modified CASL chain}
  \label{subfig: Delay nuc in CASL}
  \end{subfigure}
  \caption{We capture \protect\subref{subfig: ENDF delay nucs} the delays from decay pathways in the ENDF chain \protect\subref{subfig: Delay nuc in CASL} with a delay nuclide.}
  \label{fig: Delay nuclides}
\end{figure}
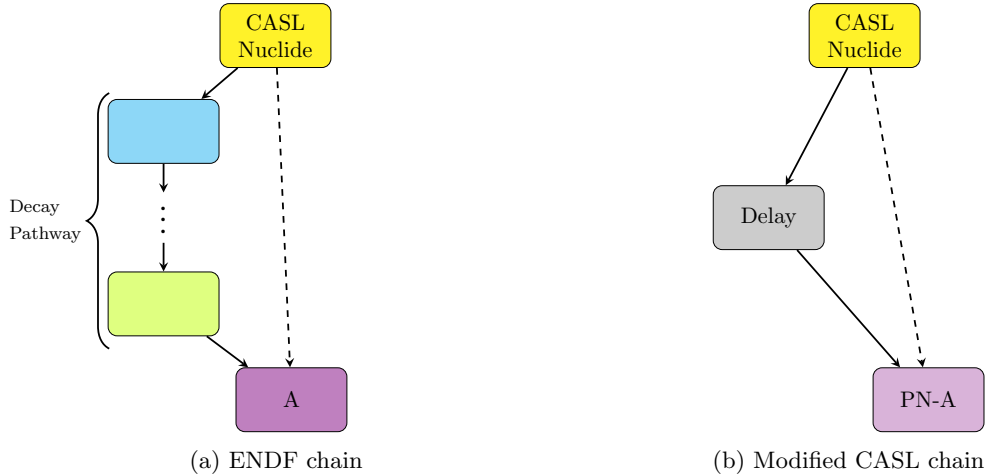

Figure~\ref{fig: Delay nuclides} demonstrates how delay nuclides are added to the CASL chain. The process is similar to how the production of PNs is done. Consider a CASL nuclide and its structure within the ENDF depletion chain, as represented by Figure~\ref{subfig: ENDF delay nucs}. In this example, nuclide A is produced directly from the CASL nuclide. However, nuclide A is also produced indirectly through the decay pathway shown. Whereas previously, both the direct and indirect production would be represented by direct production of PN-A (decay pathway assumed to be instantaneous), we now introduce a delay nuclide, as shown in Figure~\ref{subfig: Delay nuc in CASL}. The goal is to mimic the structure of the decay pathway rather than ignore it.

For the nuclides causing the largest error at startup and shutdown, we observe that there typically is one nuclide in the decay pathway with a significantly longer half-life than the rest. Thus, the delay nuclide is assigned the maximum half-life of all the nuclides in the decay pathway. We also assign a decay mode to the delay nuclide so that it produces the appropriate PN, in this case PN-A.

For a real nuclide in the ENDF chain, there are several different decay pathways, and they are all traversed using a queue structure.

\subsection{CASL+All\_PNs with Delay Nuclides}
\label{subsec:CASL+All_PNs with delay nuclides}

First, to ignore effects from grouping, we represented each nuclide with its own PN, and each delay between two nuclides was given its own delay nuclide. We only did this when the maximum half-life in the decay pathway was longer than 1 minute, treating shorter decays as essentially instantaneous. One can choose to do this or not depending on the size of timesteps chosen (see the end of Section~\ref{subsec:instantaneous decay causes startup and shutdown error} for examples on how timesteps can affect whether delays can be treated as instantaneous). Note that we refer to this depletion chain, with delay nuclides, as \textit{CASL+All\_PNs\_v2}.

\begin{figure}[h!]
\centering
\begin{subfigure}[h]{0.475\textwidth}
   \includegraphics[width=1\linewidth]{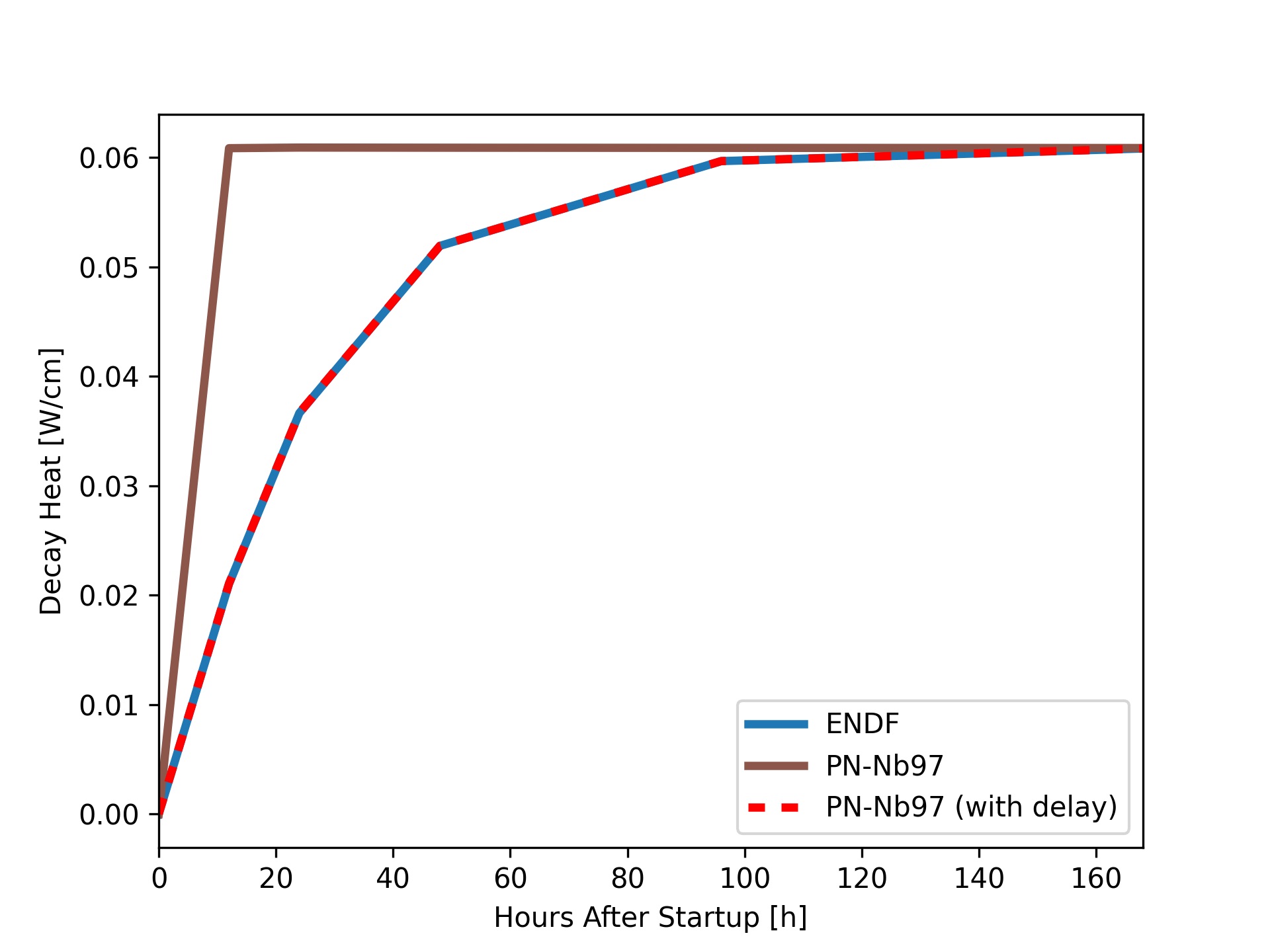}
   \caption{Startup}
   \label{subfig:Nb97_Startup} 
\end{subfigure}
\begin{subfigure}[h]{0.475\textwidth}
   \includegraphics[width=1\linewidth]{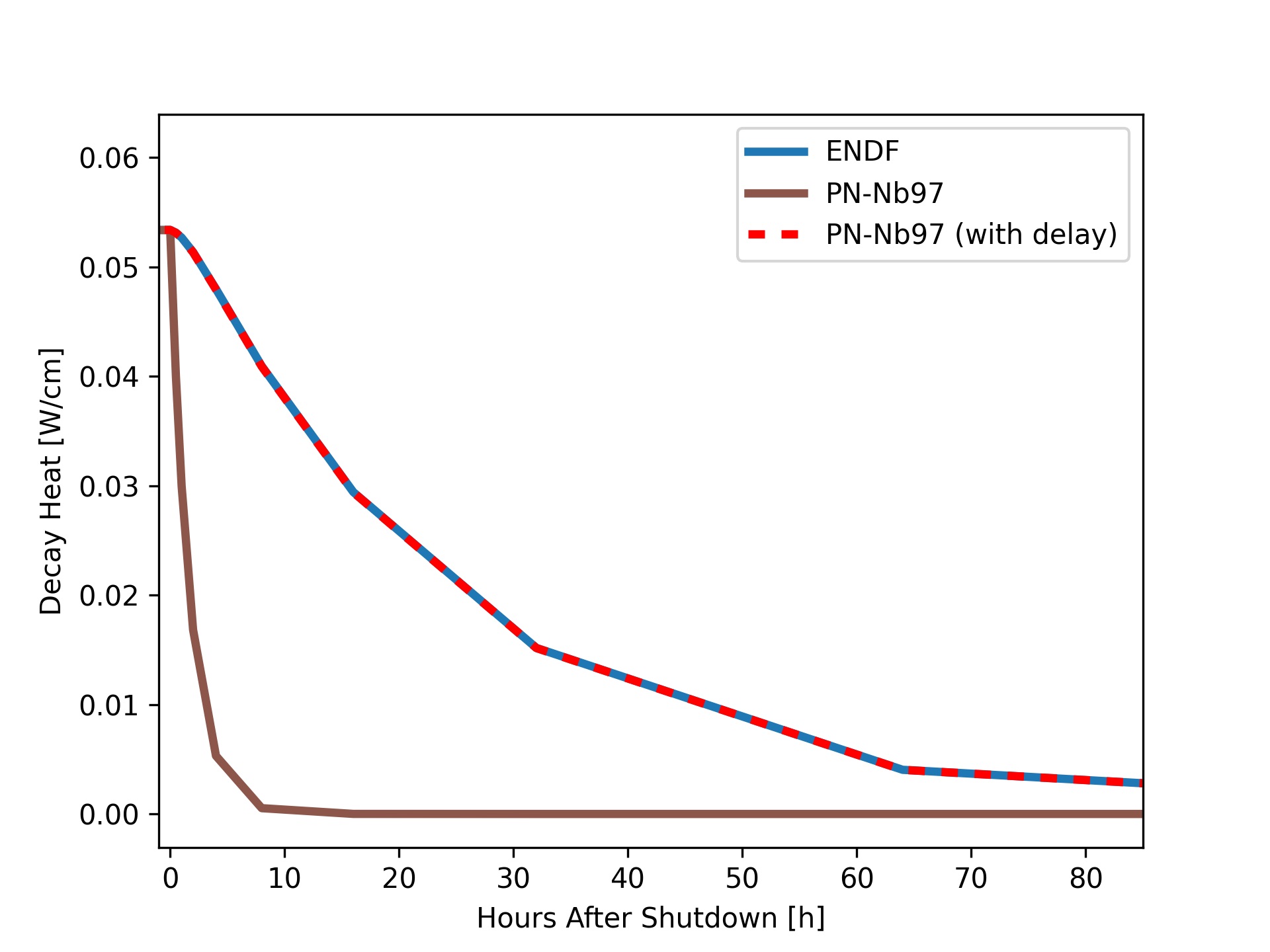}
   \caption{Shutdown}
   \label{subfig:Nb97_Shutdown}
\end{subfigure}
\caption{Decay heat of \isotope[97]{Nb} demonstrates how adding delay nuclides drastically improves decay heat both during \protect\subref{subfig:Nb97_Startup} startup and \protect\subref{subfig:Nb97_Shutdown} shutdown.}
\label{fig:Nb97_Case_Study}
\end{figure}

This chain, just like the first version, does not decrease the number of nuclides, but provides a useful starting point to see if the approach will work when we reduce the number of PNs and delay nuclides. By starting with the All\_PNs case rather than 10PNs, we examine the decay heat contributions of specific nuclides too. As an example, we consider \isotope[97]{Nb}, as shown in Figure~\ref{fig:Nb97_Case_Study}. The benchmark is the decay heat contribution of \isotope[97]{Nb} in the ENDF chain. As shown in the figure, \textit{PN-Nb97} from the CASL+All\_PNs chain (no delays) has a pretty significant error. However, adding delay nuclides results in \textit{PN-Nb97 (with delay)} from the CASL+All\_PNs\_v2 chain having much better agreement with the ENDF chain.

\begin{figure}[h!]
    \centering
    \includegraphics[width=70mm]{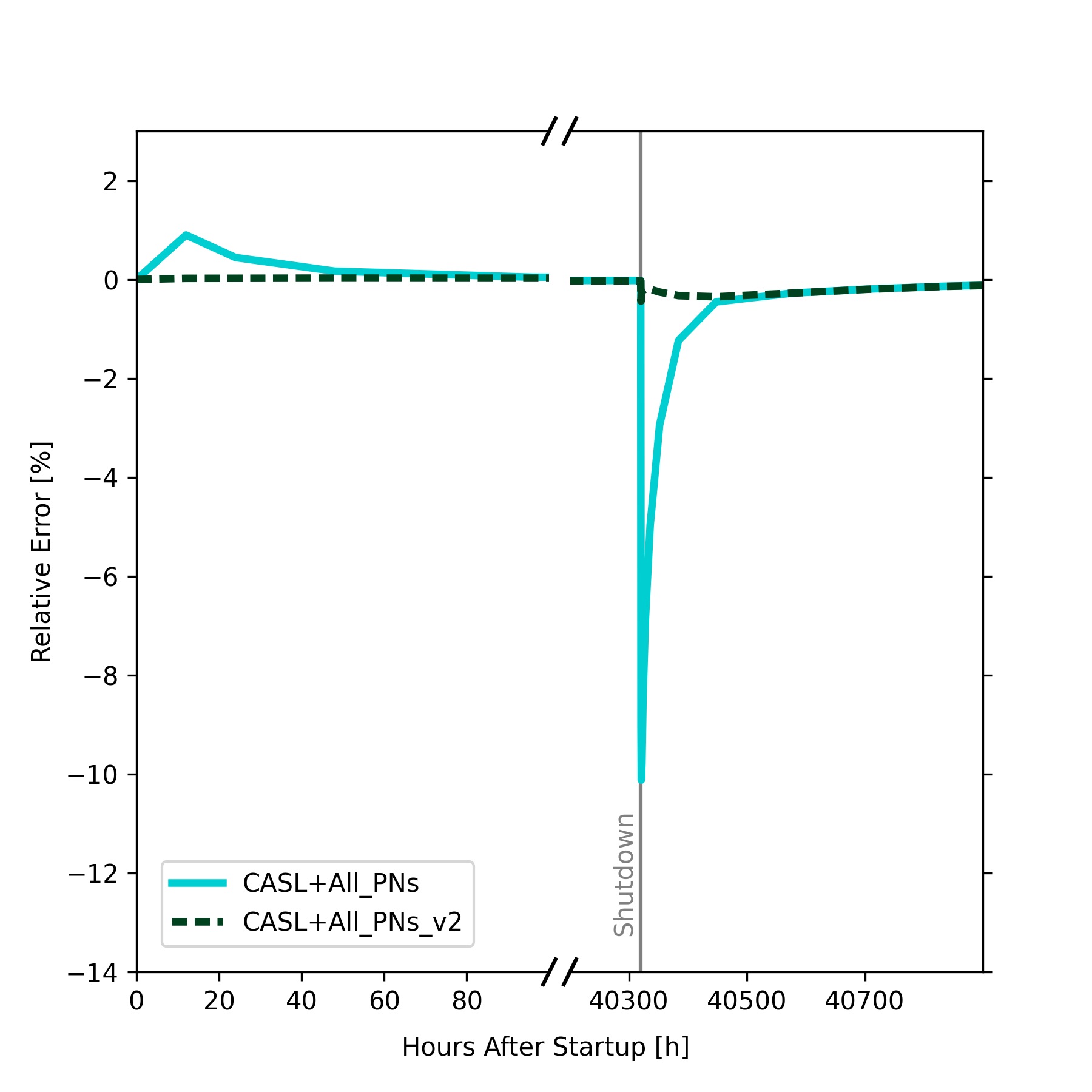}
    \caption{Error in decay heat drastically decreases for CASL+All\_PNs with the addition of delay nuclides.}
    \label{fig:All_PNs v1 vs v2}
\end{figure}

We can also consider total decay heat from the CASL+All\_PNs\_v2. As shown in Figure~\ref{fig:All_PNs v1 vs v2}, the addition of delay nuclides drastically reduces the error in decay heat compared to our ENDF benchmark. This chain has relatively low error even throughout startup and shutdown, matching ENDF much better than the version of the depletion chain without delays. The runtime performance of this chain is comparable to the CASL chain, as the majority of runtime for OpenMC comes from tallies during transport.

\subsection{CASL+10PNs with Delay Nuclides}
\label{subsec:CASL+10PNs with delay nuclides}

We now aim to use delay nuclides to similarly improve the results of the CASL+10PNs chain. Not only will we reduce the number of PNs, but we also want to decrease the number of delay nuclides, recalling that the motivation of a simplified chain is to have a reduced set of nuclides to track.

To review, there are two decay constants associated with each delay nuclide: $\lambda_{delay}$ (the decay constant of the delay nuclide itself) and $\lambda_{PN}$ (the decay constant of the PN into which the delay nuclide decays). In the All\_PNs case, for every decay pathway that has a delay, we let $\lambda_{delay}$ be the decay constant of the nuclide with the largest half-life in the pathway. We also let $\lambda_{PN}$ be the PN that was representing the individual nuclide. This resulted in several hundred different delay nuclides.

In this 10 PNs case, there are only ten options for $\lambda_{PN}$, which already decreases the number of delay nuclides. However, we also want to group the $\lambda_{delay}$'s. This helps preserve the computational benefits that come with tracking relatively few nuclides.

For the PNs, the groups remain the same as shown in Figure~\ref{fig:10_PNs}. In order to ascertain which delay nuclides to include and how to group them, we construct a histogram of important delays to capture. We conduct the following procedure. We take the 50 nuclides with the largest errors in decay heat one hour post shutdown, as the largest relative error for CASL+All\_PNs was nearly right after shutdown.

Recall that $\lambda_{delay}$ for a particular decay pathway is chosen from the nuclide with the maximum half-life (MHL) in that pathway. For each of the 50 nuclides with largest error, we look at all such nuclides in each of the possible pathways of production as we iterate through the chain. For instance, depending on the pathway producing \isotope[97]{Nb}, the MHL nuclide could be \isotope[97\text{m1}]{Nb} or \isotope[97]{Zr}. The majority of the actual error comes from the \isotope[97]{Zr} to \isotope[97]{Nb} decay since the other decay has a much shorter half-life on the order of seconds, which we can treat as instantaneous anyways. Thus, for the purposes of making the histogram, we restrict to delays with half-lives larger than 5 minutes. These are the delays that are likely actually accounting for the majority of the error.

\begin{figure}[ht]
  \centering
  \includegraphics[width=70mm]{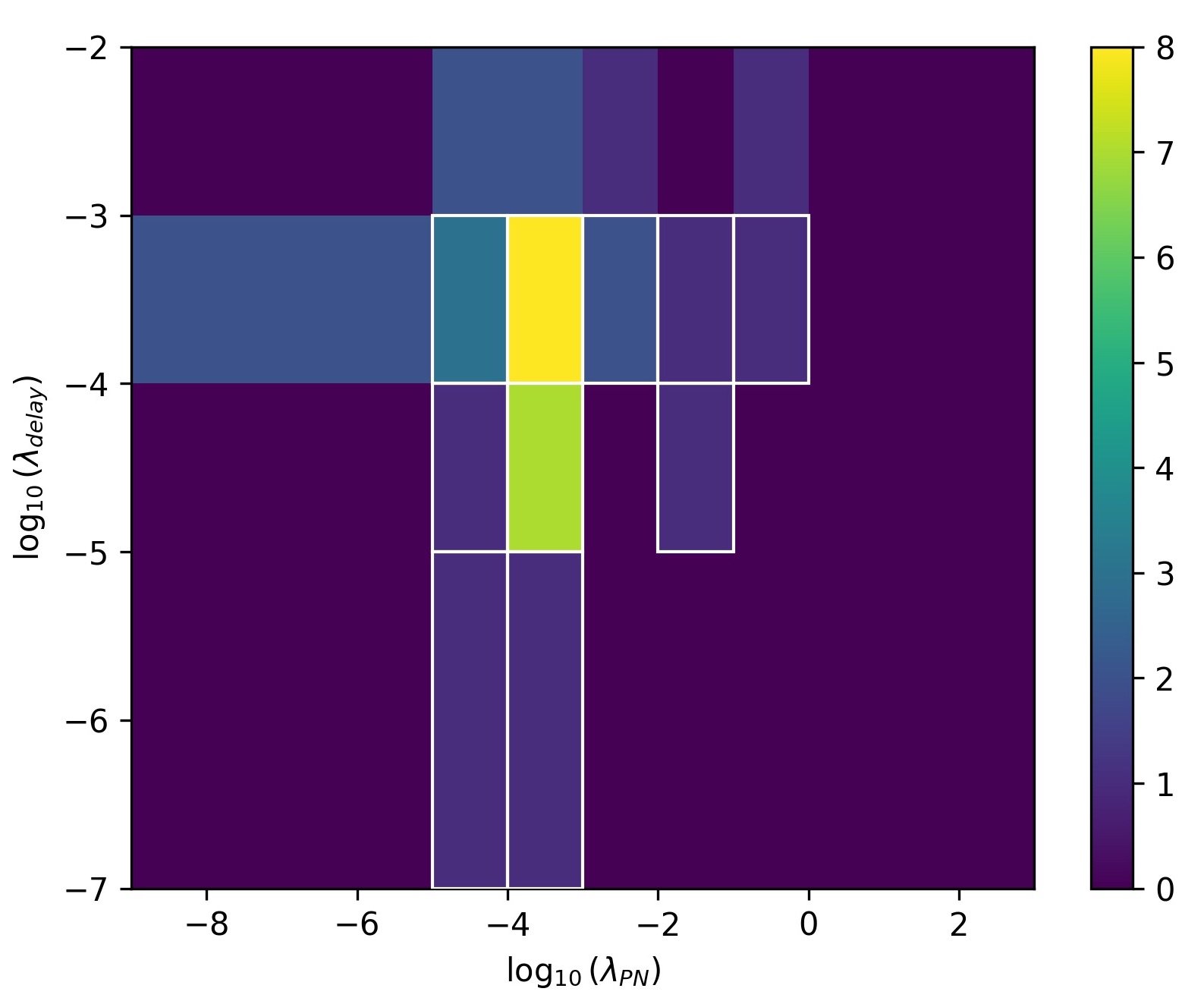}
  \caption{Histogram of important delays to capture from 50 nuclides with largest error 1 hour after shutdown.}
  \label{fig:Delay_Histogram}
\end{figure}

Figure~\ref{fig:Delay_Histogram} shows a 2D histogram of these delays, demonstrating which delay nuclides are most needed. The sections surrounded by white represent a region for which we choose to add a delay nuclide.

For example, the upper and leftmost white rectangle represents a delay with $\lambda_{delay}$ between $10^{-4}$ and $10^{-3}$ that decays to the PN representing nuclides with decay constants between $10^{-5}$ and $10^{-4}$.

Note that before introducing delay nuclides, we were essentially forcing instantaneous decay. Thus, in order to prevent erroneously forcing a long-lived decay to happen very quickly, we had imposed a cutoff of 1 week when considering decays. Only if the half-life of a decay was less than 1 week would we produce a PN. With delays, there is greater flexibility in changing this cutoff as one desires. The reason being that with delay nuclides, long-lived decays can be better represented. However, out of the 50 nuclides with largest error, only a pathway producing \isotope[115\text{m1}]{In} does not meet the 1 week cutoff, so we did not change the cutoff. Furthermore, as we only chose a relatively small number of delay nuclides, we still would like to prevent forcing instantaneous decays for the decays that are not rerouted through a delay. Keeping the same cutoff also makes comparisons to the initial versions of the chains simpler.

After deciding which delay nuclides to create based on Figure~\ref{fig:Delay_Histogram}, we had to choose a representative $\lambda_{delay}$ for each delay, just as we had chosen a representative $\lambda_{PN}$ for all the nuclides in the group of a single PN. To do this, for each delay group, we took the average of the decay constants of the MHL nuclides in decay pathways for the 50 nuclides with largest error if they corresponded to that group.

Finally, we created the delay nuclides. Then, we iterated through the ENDF depletion chain, just as we described for the addition of PNs to the CASL chain. The difference is that if the delay for a particular pathway belonged to one of the delay groups, instead of directly producing a PN, we produced a delay nuclide that eventually would decay into the appropriate PN. 

\begin{figure}[h]
  \centering
  \includegraphics[width=70mm]{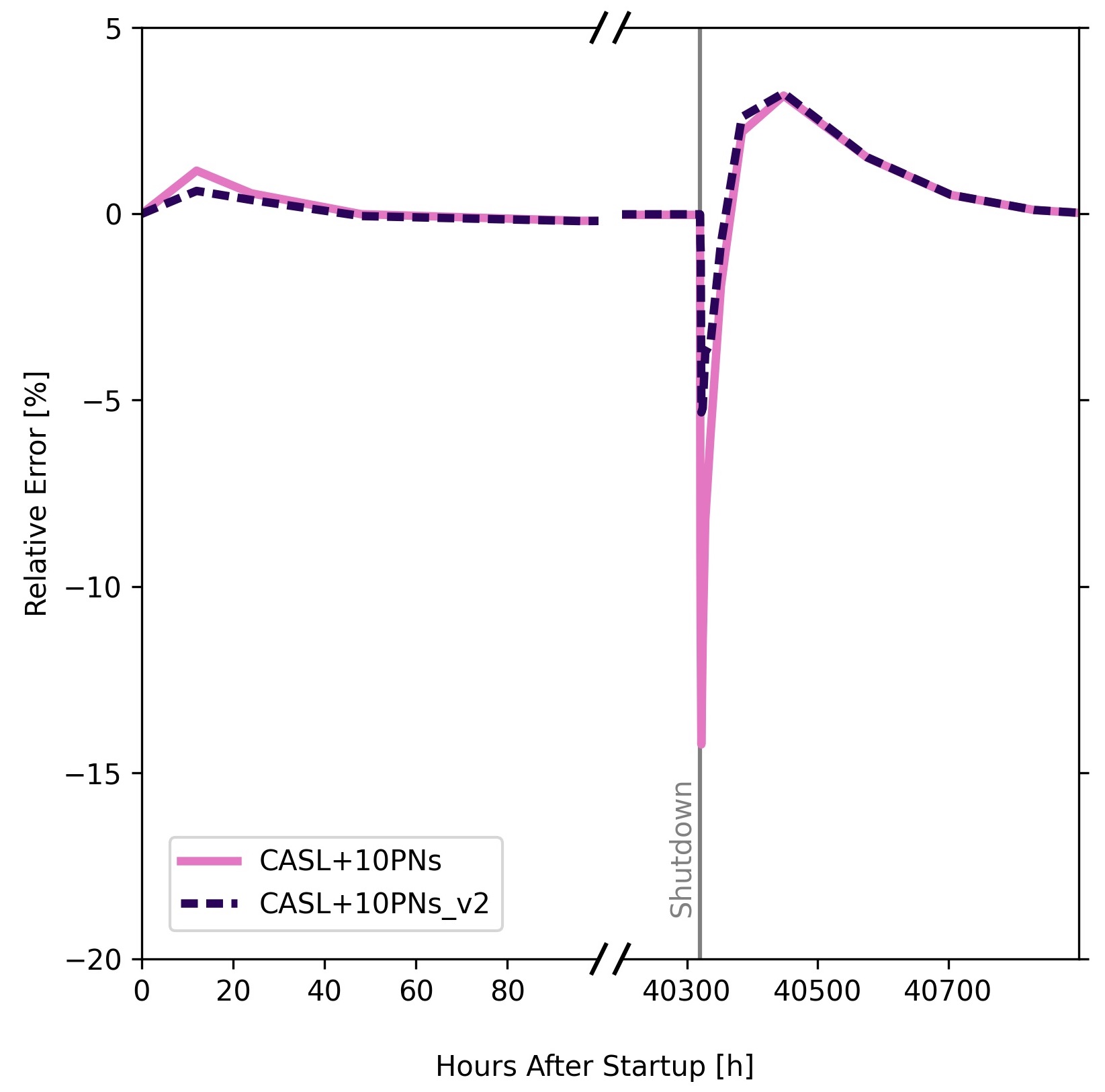}
  \caption{Error in decay heat also decreases significantly for CASL+10PNs with the addition of delay nuclides.}
  \label{fig:Relative_Error_10PNs_v1_v2}
\end{figure}

We ran the depletion problem with this new CASL+10PNs\_v2 chain. The runtime performance of this chain is also comparable to the CASL chain. The results shown in Figure~\ref{fig:Relative_Error_10PNs_v1_v2} demonstrate that the startup overestimation and shutdown underestimation errors have been reduced significantly after adding delay nuclides.

There still appear to persist some errors. For instance, the overestimation spike after the shutdown underestimation has not decreased with the addition of delay nuclides. However, we found that this error is an effect of our choice of the ten specific representative $\lambda_{PN}$'s. In fact, with slightly different choices of $\lambda_{PN}$'s (e.g., using geometric mean), this secondary spike decreases. Furthermore, these spikes do not appear in Figure~\ref{fig:All_PNs v1 vs v2}, confirming that it is an effect of representing all the nuclides with just ten groups. In other words, the delay nuclides really do an impressive job of matching the production rate.

\begin{figure}[h]
  \centering
  \includegraphics[width=70mm]{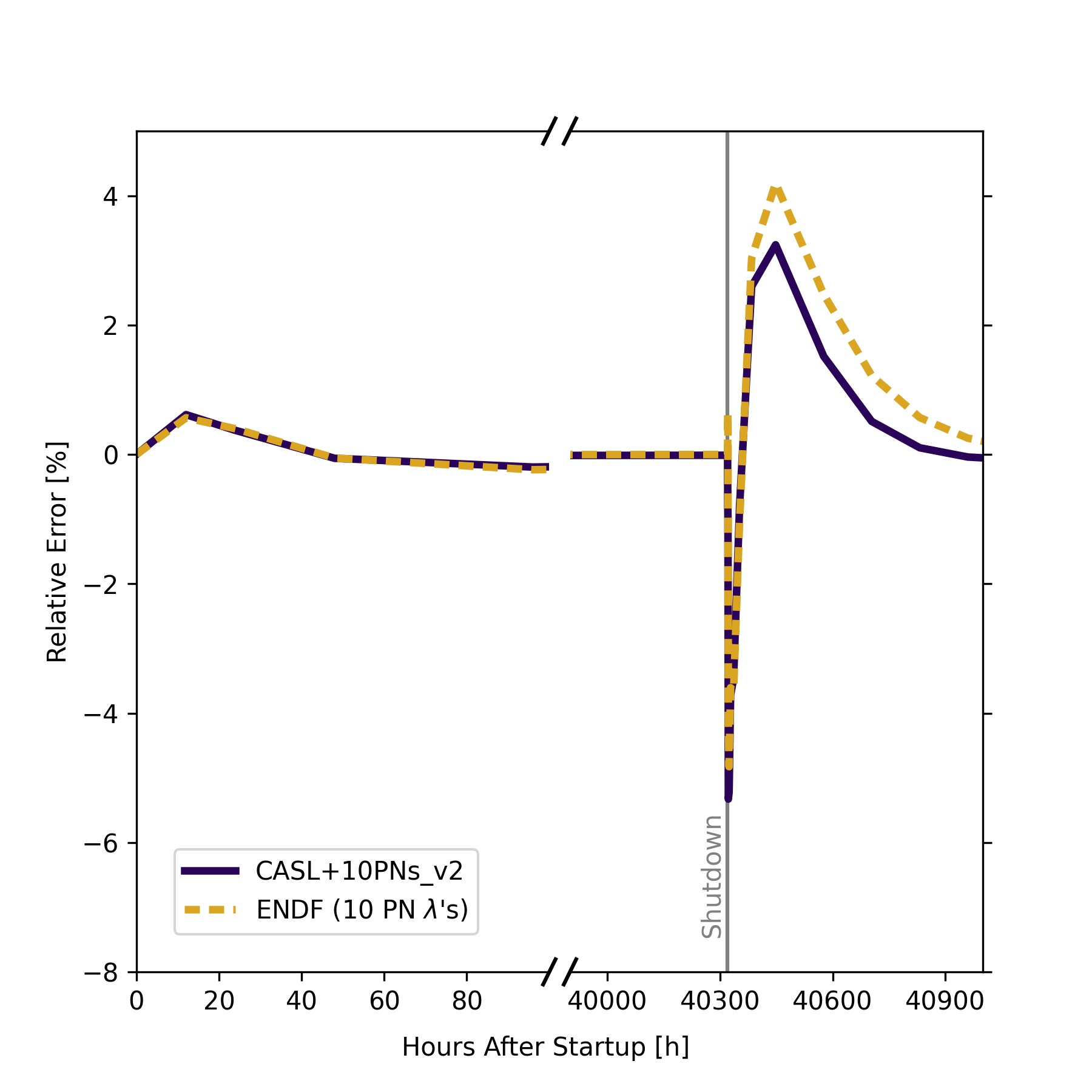}
  \caption{Adding delay nuclides reduces error nearly as much as possible given the choice of $\lambda_{PN}$'s.}
  \label{fig:Relative_Error_10PNs_v2_ENDF_Prod}
\end{figure}

There is another way for us to confirm that the delay nuclides improve production rates and reduce error nearly as well as they can with the ten groups we have created. We create a new depletion chain by taking the ENDF chain and replacing the decay constant of each ENDF-exclusive nuclide with the appropriate one of the ten representative decay constants. No decay pathways have been modified at all, since we have not actually decreased the number of nuclides. So the production rates, up to errors in decay constants, should be fairly accurate. Figure~\ref{fig:Relative_Error_10PNs_v2_ENDF_Prod} shows the error in CASL+10PNs\_v2 and this modified ENDF chain. Both have similar spikes in error, confirming that the delay nuclides do really help the production rates in the simplified chain match the production rates in the benchmark ENDF chain.

\subsection{Accuracy of Time-Integrated Decay Heat}
\label{subsec:accuracy of time-integrated decay heat}
In order to assess the accuracy of each of the chains, we also considered a slightly different metric: the time-integrated decay heat over various time intervals starting at shutdown. Given that decay heat is a measure of power, this new quantity represents the total energy released from decay after shutdown. We calculated the relative error for each chain as compared to the ENDF chain for several time intervals (Table~\ref{tab:time-integrated decay heat}). 

The CASL chain has pretty large relative error, especially in the shorter time intervals. As the interval gets longer, the short-lived nuclides that are not included in CASL tend to decay away, reducing the relative error in the CASL chain.

Adding delay nuclides decreases relative error for both the 10PNs and All\_PNs chains. The CASL+10PNs\_v2 chain remains under 5\% error for all time intervals analyzed, and the CASL+All\_PNs\_v2 chain remains under 0.3\% error.

\begin{table*}[b]
\centering
\caption{Error in Time-Integrated Decay Heat Post Shutdown}
\label{tab:time-integrated decay heat}
\begin{tabular}{@{} l *{6}{c} @{}}
\midrule
\smash[b]{\begin{tabular}[t]{@{}c@{}} Depletion\\Chain \end{tabular}} & \multicolumn{6}{c@{}}{Hours after Shutdown}\\
\cmidrule(l){2-7}
& 1 & 2 & 16 & 64 & 128 & 512   \\ 
\midrule
  CASL  & -68.46\% &  -59.71\% &  -31.33\% &  -16.56\% &  -11.28\% &  -5.51\% \\
  CASL+10PNs  & -4.20\% &  -6.84\% &  -8.49\% &  -4.16\% &  -1.58\% &  0.18\% \\
  CASL+10PNs\_v2  & -0.38\% &  -1.18\% &  -3.41\% &  -1.52\% &  0.16\% &  0.91\% \\
  CASL+All\_PNs & -4.07\% &  -5.91\% &  -6.76\% &  -4.32\% &  -3.02\% &  -1.41\% \\
  CASL+All\_PNs\_v2 & -0.19\% &  -0.22\% &  -0.21\% &  -0.24\% &  -0.28\% &  -0.26\% \\
\end{tabular}
\end{table*}

\section{Sodium-Cooled Fast Reactor}
\label{sec:sodium-cooled fast reactor}

We have now established that adding PNs and delay nuclides improves CASL decay heat estimates for the specific PWR pin cell problem we were working with. However, the process of creating the appropriate PNs and delay nuclides is general, not specific to the PWR problem. Thus, we verify that our approach is generalizable.

We worked with a sodium-cooled fast reactor (SFR) problem, using the pin cell model described earlier in Section~\ref{subsubsec:sfr depletion model}. The neutron flux spectrum is different for this case, leading to different capture branching ratios. To approximate the energy dependence of capture branching ratios, OpenMC's website provides two versions of each depletion chain, one with capture branching ratios averaged over a thermal flux spectrum, and one over a fast flux spectrum. To build our modified depletion chains for the SFR model, we used the ENDF and CASL chains from the fast spectrum.

\begin{figure}[h!]
\centering
\begin{subfigure}[h]{0.475\textwidth}
   \includegraphics[width=1\linewidth]{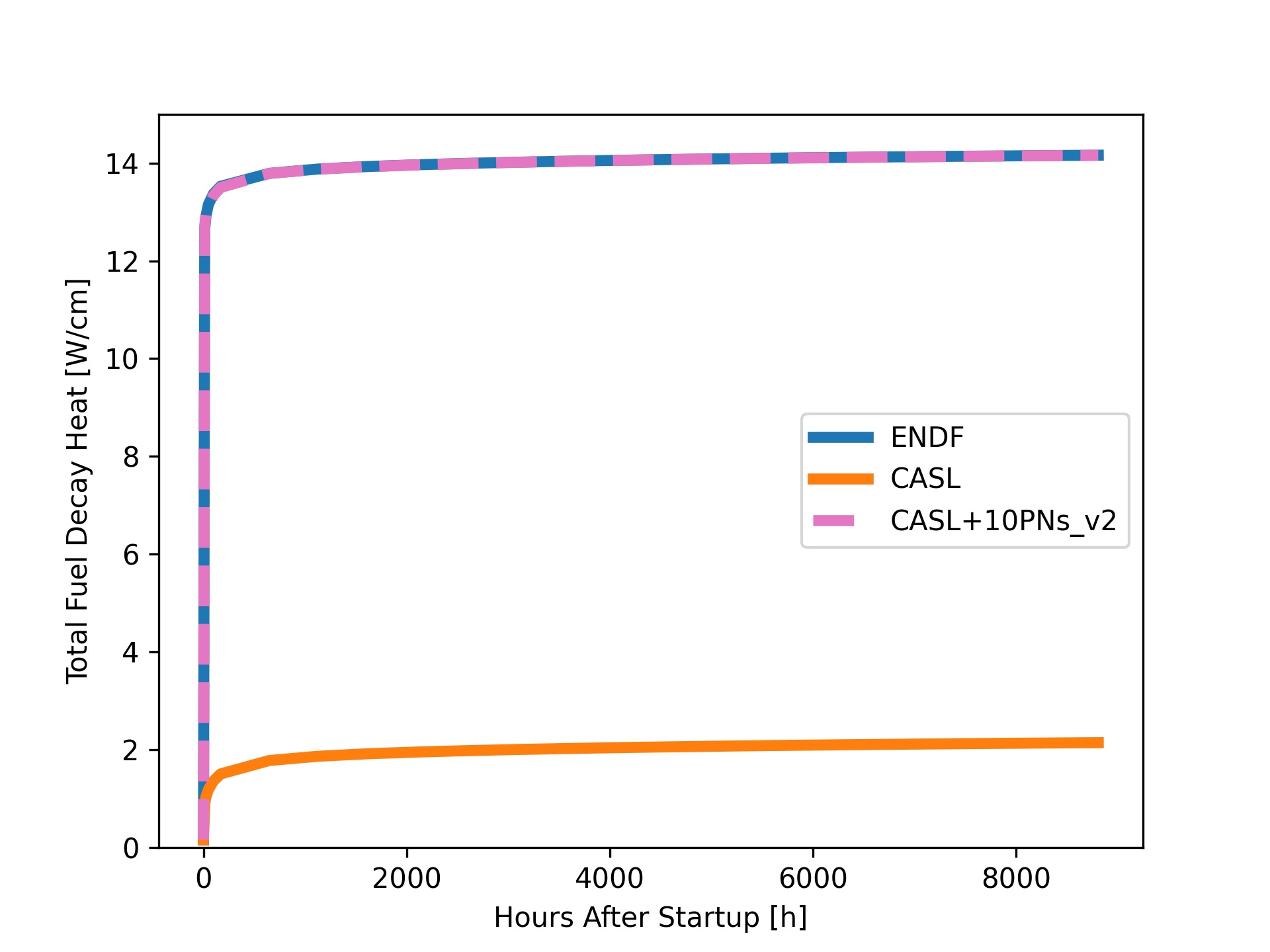}
   \caption{Operation}
   \label{fig:SFR - Decay Heat - Operation} 
   \vspace{-3pt}
\end{subfigure}
\begin{subfigure}[h]{0.475\textwidth}
   \includegraphics[width=1\linewidth]{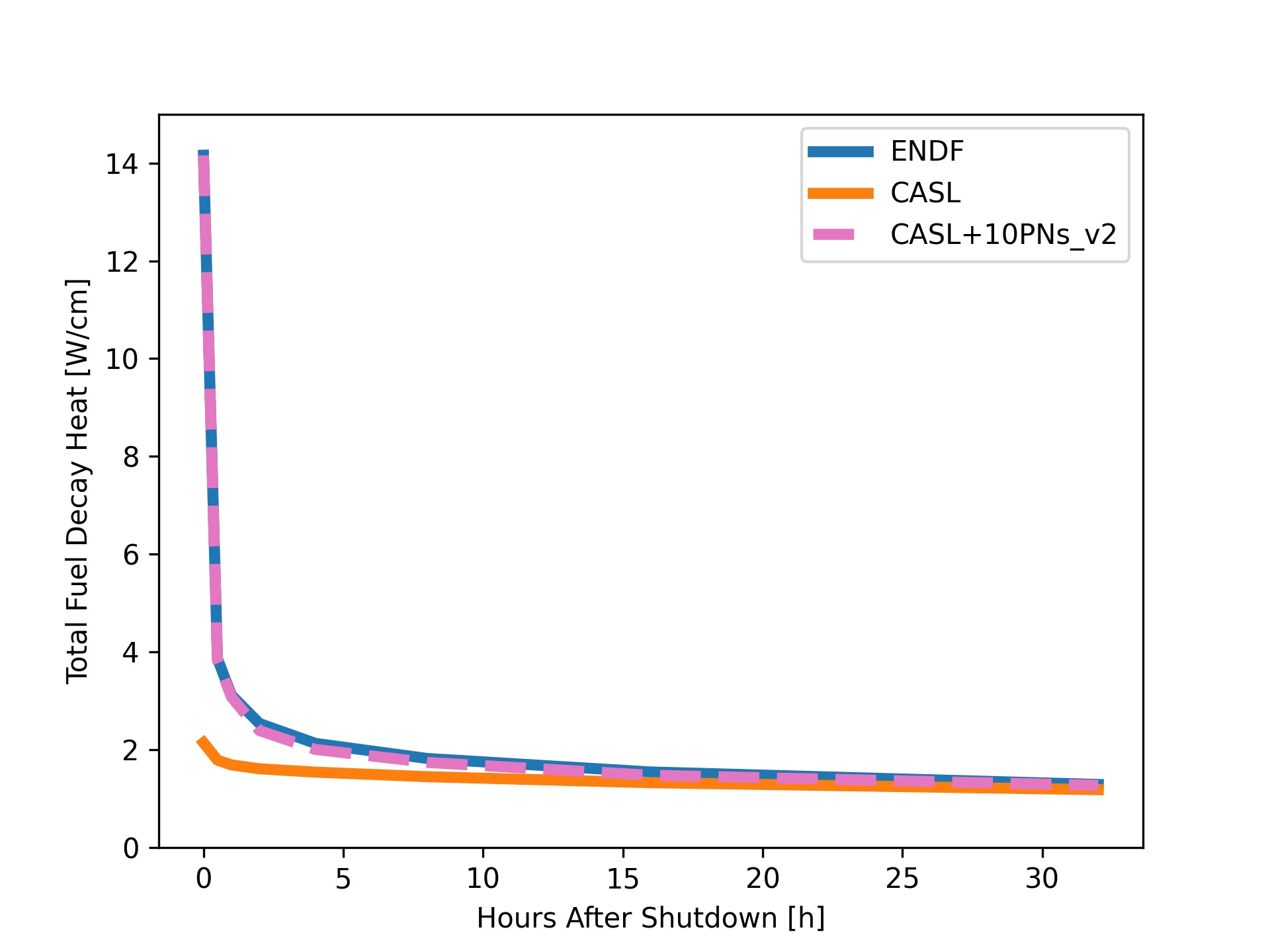}
   \caption{Post shutdown}
   \label{fig:SFR - Decay Heat - Post-shutdown}
\end{subfigure}
\caption{Even for a SFR pin cell, the CASL chain underestimates decay heat during operation and after shutdown. The modified chain with PNs and delay nuclides matches ENDF much better.}
\label{fig:SFR - Decay Heat}
\end{figure}

Figure~\ref{fig:SFR - Decay Heat} demonstrates that even for a fast spectrum, the CASL chain drastically underestimates decay heat. The cause, once again, is the cumulative effect of the many nuclides that are exclusively in the ENDF chain, and not in the CASL chain.

For this SFR model, the geometry and materials are different from the PWR. However, due to the physics-based nature of our approach for the PWR, and since nothing was fitted to our specific problem, we employed the same algorithm to create our modified depletion chains. We thus created a CASL+10PNs\_v2 chain for the SFR problem, with both 10 PNs and 10 delay nuclides. The PNs and delay nuclides have the same $\lambda_{PN}$'s and $\lambda_{delay}$'s as the PWR chain, whose values are given in \hyperref[sec:appendix A]{Appendix A}. Only some capture branching ratios producing PNs or delay nuclides are different between the PWR and SFR chains, inherited from the differences in capture branching ratios in the thermal and fast spectrum ENDF and CASL chains. As shown in Figure~\ref{fig:SFR - Decay Heat}, the decay heat was captured much more accurately with this modified chain.

\begin{figure}[h!]
\centering
\begin{subfigure}[h]{0.47\textwidth}
   \includegraphics[width=1\linewidth]{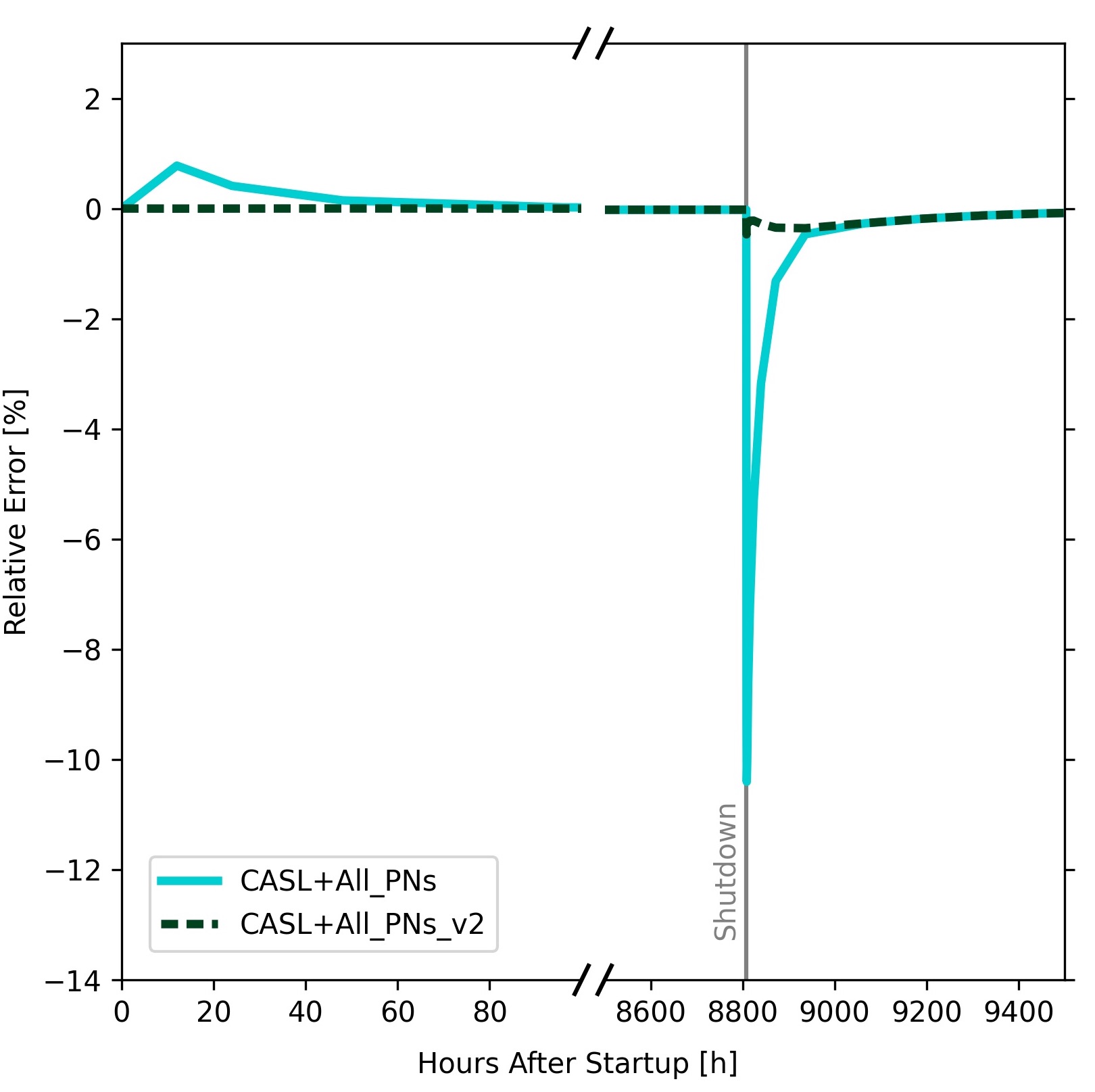}
   \caption{CASL+All\_PNs}
   \label{subfig:SFR_Relative_Error_All_PNs_v1_v2} 
\end{subfigure}
\begin{subfigure}[h]{0.47\textwidth}
   \includegraphics[width=1\linewidth]{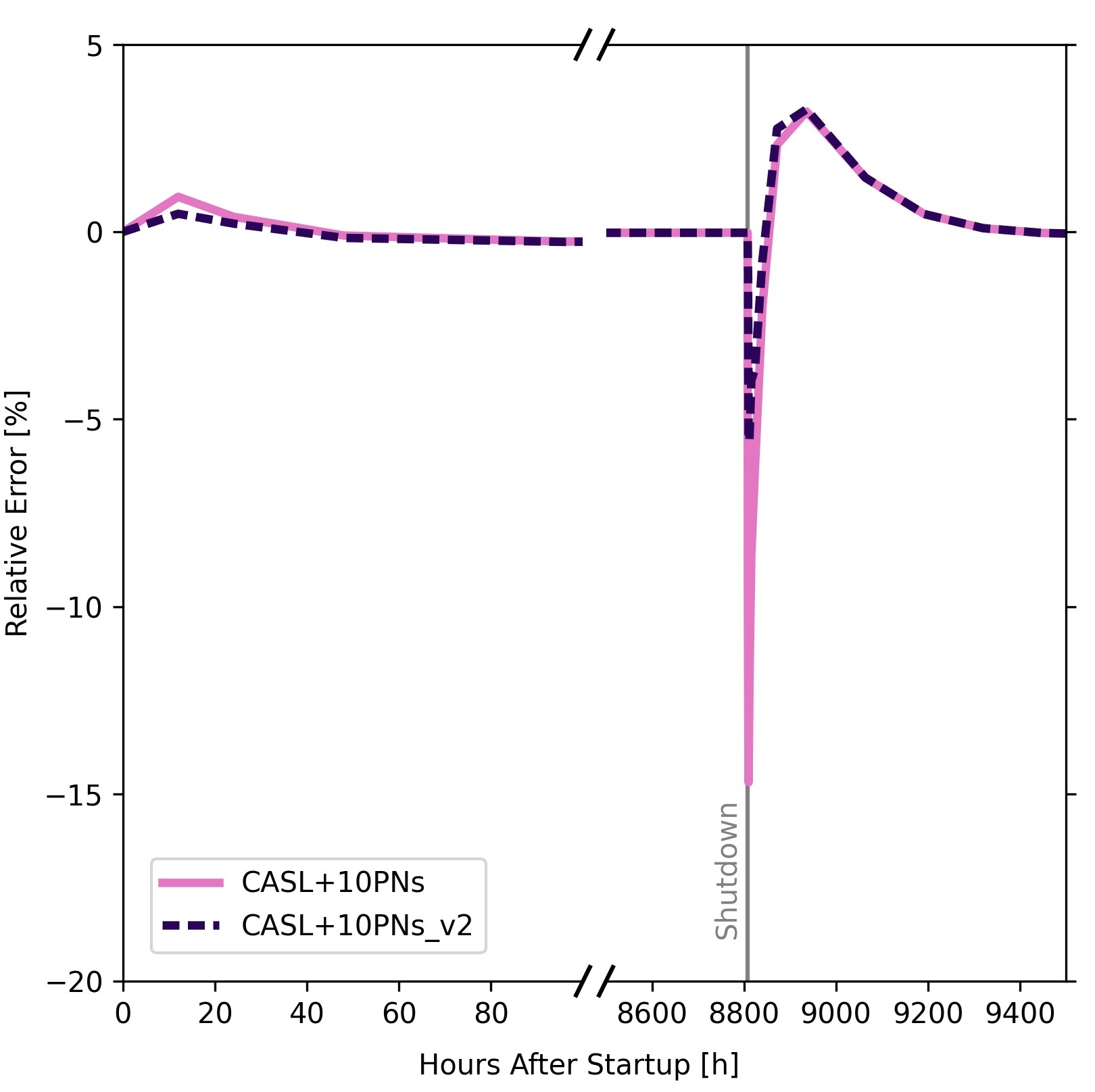}
   \caption{CASL+10PNs}
   \label{subfig:SFR_Relative_Error_10PNs_v1_v2}
\end{subfigure}
\caption{For SFR model, the addition of delay nuclides also reduces error for both \protect\subref{subfig:SFR_Relative_Error_All_PNs_v1_v2} CASL+All\_PNs and \protect\subref{subfig:SFR_Relative_Error_10PNs_v1_v2}  CASL+10PNs.}
\label{fig:SFR_Relative_Error_Both_v1_v2}
\end{figure}

We created CASL+All\_PNs and CASL+All\_PNs\_v2 for the SFR case too, and demonstrated that delay nuclides improve production rate accuracies, resulting in decreased errors (Figure~\ref{subfig:SFR_Relative_Error_All_PNs_v1_v2}). Furthermore, we used the same procedure as in the PWR case to add 10 PNs to the CASL chain, creating a CASL+10PNs for the SFR problem. Via a similar procedure to the PWR case, we added 10 delay nuclides, resulting in the CASL+10PNs\_v2 chain. The relative errors in these two chains show similar patterns to the PWR problem, with the addition of delay nuclides significantly reducing the spikes in relative error that originate from instantaneous decays (Figure~\ref{subfig:SFR_Relative_Error_10PNs_v1_v2}). A consideration of time-integrated decay heat after shutdown also showed similar results to the PWR case.

\section{Conclusion}
\label{sec:conclusion}

In this work, we use OpenMC to run depletion simulations, from which we calculate decay heat. The CASL simplified depletion chain tracks around an order of magnitude fewer nuclides than the ENDF depletion chain, providing runtime and memory benefits. However, the CASL chain drastically underestimates decay heat when comapared to decay heat from simulations that use the ENDF chain, our benchmark.

We identified that this discrepancy is due to the cumulative decay heat of the many nuclides not included in the CASL chain. We utilized pseudo-nuclides (PNs) to more accurately model decay heat without sacrificing the computational benefits of the CASL chain. We present the algorithm we used to add 10 PNs to the CASL chain,  creating the CASL+10PNs chain, and resulting in improved decay heat accuracy.

We observed that using PNs alone in our algorithm, even with one PN per nuclide  in the CASL+All\_PNs chain, there remained a significant error during startup and shutdown. Our analysis showed that this was due to mistimed production of PNs. We introduced delay nuclides, which provide enhanced accuracy by better reflecting the structure of the decay pathways present in the full ENDF chain. By adding delay nuclides to CASL+All\_PNs and adding 10 delay nuclides to CASL+10PNs, we obtained the CASL+All\_PNs\_v2 and CASL+10PNs\_v2 chains, respectively.

For OpenMC, reaction tallies are the main cause for runtime increase. Thus, if decreasing runtime is the only goal, one can use the CASL\_All\_PNs\_v2 chain. Alternatively, one can remove negligible reactions from the ENDF chain and use that instead, which may not improve runtime as much, but will provide a better representation of the decay chain structure. All of the simplified depletion chains created by adding relatively few PNs or delay nuclides ran around 50\% faster in terms of total runtime. When only considering depletion solver time, the number of nuclides became more important, with the 10PNs chains showing better runtime improvement than the All\_PNs chains.

After adding the delay nuclides, we drastically improved production rate accuracy. For the PWR model, relative error in time-integrated decay heat for CASL+All\_PNs\_v2 remained below 0.3\% for the time intervals considered. Furthermore, the CASL+10PNs\_v2 chain remained below 5\% relative error. Overall, the implementation of delay nuclides proposed in this paper drastically improves accuracy in decay heat estimates.

Future work will explore different methods of determining the PNs. For instance, there is freedom in how the boundaries between the groups themselves is determined. Furthermore, one can investigate the choice of representative decay constants within each group. For instance, $\lambda_{PN}$ could be chosen by taking the average of the logarithm of the decay constants of nuclides in the group. Note that this is equivalent to taking the geometric mean of the decay constants in a group. Preliminary testing showed that this approach had less absolute startup and shutdown error when used with 10 PNs. At the same time, the error appeared to oscillate more. For example, a large underestimation after shutdown would result in a small overestimation before settling to zero error. One can also optimize PNs for a representative problem, such as an infinite lattice, and then utilize the chain for a larger, more time-intensive simulation~\cite{Griesheimer_Large_Scale_MC}.

The use of PNs and delay nuclides opens up possibilities for improving simplified models of processes involving decay pathways. It is especially applicable when not all nuclides need to be tracked individually, but the overall process depends on the contribution of many different nuclides.

\pagebreak
\section{Appendix A: Pseudo-Nuclides and Delay Nuclides}
\label{sec:appendix A}

The solid black lines in Figure~\ref{fig:10_PNs} demarcate the boundaries between the ten groups that we represent with PNs. The boundaries are at $10^{-9}, 10^{-5}, 10^{-4}, 10^{-3}, 10^{-2}, 10^{-1}, 10^0, 10^1,$ and $10^3 \; s^{-1}.$ By taking the mean of decay constants of nuclides within a group, we chose a representative $\lambda_{PN}$ to assign to the PN associated with that group. The decay constants of each of the PNs used in the CASL+10PNs and CASL+10PNs\_v2 chain are shown in Table~\ref{tab:appendix PNs}.

\begin{table*}[h]
\centering
\caption{Decay Constants of PNs}
\label{tab:appendix PNs}
\begin{tabular}{@{} l *{10}{c} @{}}
& PN-0 & PN-1 & PN-2 & PN-3 & PN-4 \\
\midrule
  $\lambda_{PN} \; [s^{-1}]$ & $8.83\times 10^9$ &  $3.03\times 10^5$ & $1.57 \times 10^4$ & $1.74 \times 10^3$ & $1.89\times 10^2$\\
  \\ \\
  & PN-5 & PN-6 & PN-7 & PN-8 & PN-9 \\
  \midrule
  $\lambda_{PN} \; [s^{-1}]$ & $1.88\times 10^1$ &  1.65 &  $1.98\times 10^{-1}$ & $1.87\times 10^{-2}$ &  $7.45\times 10^{-7}$ \\
\end{tabular}
\end{table*}
\bigskip

For the CASL+10PNs\_v2 chain, in addition to PNs, we add 10 delay nuclides, as discussed in Section~\ref{sec:delay nuclides}. The delay nuclides decay to PNs with their own decay constants. Delay nuclides starting with ``1$\rightarrow$'' capture decay pathways in which the maximum half life nuclide has a decay constant of $10^{-7}$ to $10^{-5}\;s^{-1}.$ Delay nuclides starting with ``2$\rightarrow$'' capture decay pathways in which the maximum half life nuclide has a decay constant of $10^{-5}$ to $10^{-4}\;s^{-1}.$ Delay nuclides starting with ``3$\rightarrow$'' capture decay pathways in which the maximum half life nuclide has a decay constant of $10^{-4}$ to $10^{-3}\;s^{-1}.$ The decay constants of each of the delay nuclides was chosen based on the procedure described in Section~\ref{subsec:CASL+10PNs with delay nuclides}, and the decay constants are shown in Table~\ref{tab:appendix Delays}.
\begin{table*}[h]
\centering
\caption{Decay Constants of Delay Nuclides}
\label{tab:appendix Delays}
\begin{tabular}{@{} l *{10}{c} @{}}
& 1$\rightarrow$PN-2 & 1$\rightarrow$PN-3 & 2$\rightarrow$PN-2 & 2$\rightarrow$PN-3 & 2$\rightarrow$PN-5 \\
\midrule
  $\lambda_{delay} \; [s^{-1}]$ & $7.57\times 10^4$ & $1.20\times 10^5$ & $9.76 \times 10^3$ & $1.29 \times 10^4$ & $6.03 \times 10^4$\\
  \\ \\
  & 3$\rightarrow$PN-2 & 3$\rightarrow$PN-3 & 3$\rightarrow$PN-4 & 3$\rightarrow$PN-5 & 3$\rightarrow$PN-6 \\
  \midrule
  $\lambda_{delay} \; [s^{-1}]$ & $1.15\times 10^3$ &  $1.46\times 10^3$ &  $1.34\times 10^3$ & $3.32\times 10^3$ &  $3.02\times 10^3$ \\
\end{tabular}
\end{table*}

The yields and branching ratios of both PNs and delay nuclides are computed from the algorithm described in Section~\ref{subsubsec:implementation of PNs}, which uses data from the full ENDF chain.

\pagebreak
\section*{Acknowledgments}

This research was supported by the Columbia University Science Research Fellows Program. The authors would also like to thank Dr. Gavin Ridley for proposing this research topic. Furthermore, the authors acknowledge the MIT Office of Research Computing and Data for providing high-performance computing resources that have contributed to the research results reported within this paper.

\pagebreak
\bibliographystyle{style/ans_js}
\bibliography{bibliography}

\end{document}